\documentclass[%
 aip,
 amsmath,amssymb,
 reprint,%
]{revtex4-1}

\usepackage{graphicx}% Include figure files
\usepackage{dcolumn}% Align table columns on decimal point
\usepackage{bm}% bold math

\usepackage[utf8]{inputenc}
\usepackage[T1]{fontenc}
\usepackage{mathptmx}
\usepackage{xcolor}

\def\eg{{\it e.g.}}
\def\ie{{\it i.e.}}

\DeclareMathOperator{\csch}{csch}

\begin{document}

\preprint{}

\title{Lagrangian pair dispersion in upper-ocean turbulence\\in the presence of mixed-layer instabilities}

\author{Stefano Berti}
\thanks{Email: stefano.berti@polytech-lille.fr}
\affiliation{Univ. Lille, ULR 7512 - Unit\'e de M\'ecanique de Lille Joseph Boussinesq (UML), F-59000 Lille, France}

\author{Guillaume Lapeyre}
\affiliation{LMD/IPSL, CNRS, \'Ecole Normale Sup\'erieure, PSL Research University, 75005 Paris, France}


%%%%%%%%%%%%%%%%%%%%%%%%%%%%%%%%%%%%%%%%%%%%
\begin{abstract}
Turbulence in the upper ocean in the submesoscale range (scales smaller than the deformation radius) plays an important role 
for the heat exchange with the atmosphere and for oceanic biogeochemistry. Its dynamics should strongly depend on the seasonal cycle 
and the associated mixed-layer instabilities. The latter are particularly relevant in winter and are responsible for the formation 
of energetic small scales that extend over the whole depth of the mixed layer. The knowledge of the transport properties of oceanic 
flows at depth, which is essential to understand the coupling between surface and interior dynamics, however, is still limited. 
By means of numerical simulations, we explore the Lagrangian dispersion properties of turbulent flows in a quasi-geostrophic model 
system allowing for both thermocline and mixed-layer instabilities. 
The results indicate that, when mixed-layer instabilities are present, the dispersion regime is local 
from the surface down to depths comparable with that of the interface with the thermocline,  
while in their absence dispersion quickly becomes nonlocal versus depth. 
We then identify the origin of such behavior in the existence of fine-scale energetic structures due to mixed-layer instabilities. 
We further discuss the effect of vertical shear on the Lagrangian particle spreading and address the correlation 
between the dispersion properties at the surface and at depth, which is relevant to assess the possibility 
of inferring the dynamical features of deeper flows from the more accessible surface ones.
\end{abstract}
%%%%%%%%%%%%%%%%%%%%%%%%%%%%%%%%%%%%%%%%%%%%
%\pacs{}% insert suggested PACS numbers in braces on next line

\maketitle

%%%%%%%%%%%%%%%%%%%%%%%%%%%%%%%%%%%%%%%%%%%%
%\linenumbers\relax % Commence numbering lines
\section{\label{sec:intro} Introduction}

Oceanic motions at scales larger than few tens of km are quasi-horizontal due to the pronounced stratification 
of seawater and Earth's rotation 
and are characterized by quasi-two-dimensional turbulence. At scales around $300$~km (in the mesoscale range), 
coherent structures (almost circular vortices) 
with depths reaching $1000$~m contain most of the kinetic energy in the ocean. 
At scales around $10$~km (\ie~in the submesoscale
range) and over the water column
the flow is populated by smaller eddies and filamentary structures associated with strong gradients of physical properties (such as temperature), 
which play an important role in both physical and biogeochemical
budgets~\cite{McWilliams2016,Levy2008,Ferrari2011,Su_etal_2018,Bracco_etal_2019,Siegelman2020}. 
Such small scales are found mainly in the mixed layer~\cite{McWilliams2016} 
(the first $\approx 100$~m below the surface), 
a weakly stratified layer lying on top of a more stratified one 
known as the thermocline. 

Two mechanisms leading to the generation of these fine scales have been proposed. On one side, they can be produced by the stirring 
due to larger scale eddies~\cite{LK2006,Klein_etal_2011,RMCM2012,Capet_etal_2016}. In such a case, however, they are confined close to the surface. On the other side, they can 
result from mixed-layer instabilities, in which case they extend all over the mixed layer~\cite{BFF2007}. 
When the latter is sufficiently deep, as is the case in winter, the potential energy contained in surface buoyancy gradients 
at mesoscales, can give rise to baroclinically unstable modes with horizontal scale of $O(1-10)$~km that grow over time scales 
of $O(1)$~day. 
It has to be noted that the first mechanism does not depend on the mixed-layer depth, which strongly differs from one season to another. 
As such, it cannot account for seasonal variations of the intensity of turbulence (at small scales), which has been observed to be 
a distinctive feature of submesoscale flows~\cite{SKQS2014,CFKG2015,Qiu_etal_2018}. 

In order to explore the impact of mixed-layer instabilities on submesoscale turbulence, an attractive  quasi-geostrophic (QG) 
model was recently proposed~\cite{CFFF2016}. It describes the dynamics of two coupled fluid layers having different 
stratification properties giving rise to both mixed-layer and thermocline instabilities, thus permitting a comparison of the 
two mechanisms mentioned above. 
In the absence of a mixed layer, at sufficiently small scales the model essentially gives surface quasi-geostrophic (SQG) 
dynamics~\cite{Held_etal_1995,Lapeyre2017}, which are considered a paradigm of mesoscale-driven submesoscale generation. 
It should be noted, however that in this case submesoscales are trapped at the surface.  
As shown in Ref.~\onlinecite{CFFF2016}, by including baroclinic mixed-layer instabilities, the model 
gives rise to turbulent flows characterized by energetic submesoscales
down to the thermocline, which positively compare with 
those observed in the field in winter. 

In this work we adopt the above QG model to carry out numerical simulations resolving both the mesoscale and submesoscale ranges, 
in realistic conditions for the winter midlatitude ocean. Our main goal is to investigate the role of mixed-layer instabilities 
on the spreading process of Lagrangian tracer particles. 
Here, we ignore non-geostrophic motions such as inertia-gravity waves that act at small scales 
and at high frequencies, having also some effect on Lagrangian dispersion~\cite{Sinha_etal_2019}. 

Lagrangian statistics allow access to the stirring operated by turbulent flows, which plays an essential role 
in transport processes (\eg~of biogeochemical tracers), as well as for surface energy and heat exchanges, 
at different scales~\cite{LaCasce2008,vanSebille_etal_2018}. 
Several previous studies have addressed the relative dispersion of pairs of 
surface drifters from experimental data, and its relation with the statistical properties of the underlying 
turbulent flows~(see, \eg, Refs.~\onlinecite{LaCasce2008,LE2010,Poje_etal_2014,Poje_etal_2017,CLPSZ2017,Essink_etal_2019}),  
but the results vary from region to region and are not always conclusive about dispersion regimes. 
Interestingly, however, several evidences of enhanced dispersion at submesoscales have been recently 
provided~\cite{LE2010,BDLV2011,Schroeder_etal_2012,Poje_etal_2014,CLPSZ2017}, 
which ask for a more detailed understanding of the physical processes acting at these scales.  

Below the surface, the knowledge of flow properties is still limited, due to the complexity of performing measurements at depth. 
In this respect, Lagrangian approaches can reveal a useful tool to understand the coupling between the surface and interior dynamics. 
Not many studies of relative dispersion at depth, from float trajectories, at small temporal and spatial scales are available.  
There is, however, some evidence, at rather large depths in the western Atlantic, of dispersion being local~\cite{LB2000,OGD2005}, 
meaning governed by eddies of the same size as the pair separation distance, at scales between some tens and some hundreds 
of km, or controlled by mean shear~\cite{LB2000} (up to $100$~km). 
Nonlocal dispersion, \ie~mainly due to the largest eddies, was detected in the same area~\cite{OGD2005} at 
scales smaller than $40$~km and, more recently, in the Antarctic Circumpolar Current at depths between $500$ and $2000$~m, 
in a study resolving the $1-100$~km scale range~\cite{BLSF2019}. 

Learning, from observations or numerical simulations, how submesoscale turbulence affects the spreading of Lagrangian 
particles at different depths seems appealing also in view of the high-resolution velocity data expected from future 
satellite altimetry, as the SWOT mission~\cite{Morrow_etal_2019}. 
As Lagrangian statistics reflect Eulerian ones, such as energy spectra (see, \eg, Refs.~\onlinecite{BBRS1990,FBPL2017,Malik2018,Malik2019}), 
they may serve to assess the range of validity, in terms of spatial scales, of the satellite-derived flow. 
Furthermore, the characterization of dispersion properties below the surface can be informative about the possibility 
to extrapolate information from the surface to depth.

In our numerical study, we examine particle-pair separation statistics at 
different depths, relying on both fixed-time and fixed-scale indicators. 
From a methodological point of view, our approach shares some similarity with that of Ref.~\onlinecite{KBMP2009}, where however the focus 
was not on mixed-layer instabilities, and with that of Ref.~\onlinecite{Ozgokmen_etal_2012}, where the effect of the latter was mainly 
considered for its signature on dispersion close to the surface. 
By contrasting the results obtained with different turbulent flow dynamics, \ie~generated by thermocline-only or 
mixed-layer only instabilities, or both, we aim at identifying the resulting dispersion regimes, with an emphasis 
on their general features. Furthermore, by studying the correlation of dispersion properties at the surface and 
in deeper layers, we address the question of inferring the dynamical features of deeper flows from the more accessible surface ones.  

This article is organized as follows. The model adopted for the turbulent dynamics is presented in Sec.~\ref{sec:model}; 
the main statistical properties of the turbulent flows are illustrated in Sec.~\ref{sec:turb_flow}. The results 
of the analysis of Lagrangian pair separation are reported in Sec.~\ref{sec:disp}, where we separately focus on horizontal dispersion 
at different depths (Sec.~\ref{sec:disp_h}) and on the correlation of its properties along the vertical (Sec.~\ref{sec:fsle-ii}). 
Finally, discussions and conclusions are presented in Sec.~\ref{sec:concl}.  

%%%%%%%%%%%%%%%%%%%%%%%%%%%%%%%%%%%%%%%%%%%%
\section{\label{sec:model} Model}
\begin{figure}[h]
%  \centering
  \includegraphics[width=0.9\columnwidth]{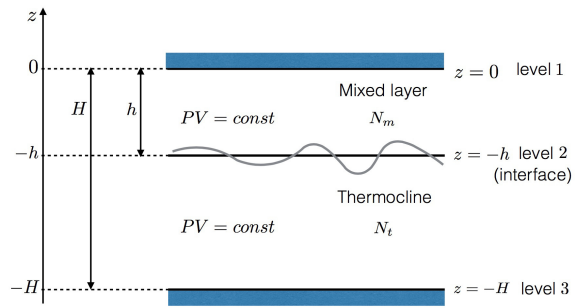}
  \caption{\label{fig:f1} Schematics of the 2-layer model.}
\end{figure}

We consider a QG model (Fig.~\ref{fig:f1}) consisting in two coupled fluid layers (aimed to represent the mixed layer and the thermocline) 
with different stratification. Such a model can give rise to both meso and submesoscale instabilities and subsequent non-linear 
turbulent dynamics that compare well with observations of wintertime submesoscale flows~\cite{CFFF2016}.
While the former instabilities are due to classical baroclinic instability, the latter are associated with mixed-layer
instabilities.
%%%%%%%%%%%%%%%%%%%%%%%%%%%%
% Fig. model schematics 
% z=0, -h, -H
% TC, ML, F 
%%%%%%%%%%%%%%%%%%%%%%%%%%%%

The model dynamics are specified by the following evolution equations (more details in Ref.~\onlinecite{CFFF2016}):
\begin{equation}
\partial_t \theta_i + J(\psi_i,\theta_i) + U_i \partial_x \theta_i + 
\Gamma_i \partial_x \psi_i = r \nabla^{-2} \theta_i + D_s(\theta_i),
\label{eq:model}
\end{equation}
where $J$ is the Jacobian operator. The fields $\theta_i$ (with $i=1,2,3$) are
three $\delta-$PV (potential vorticity) sheets at the ocean surface ($z=0$), at the base of
the mixed layer ($z=-h$) and at the bottom of the thermocline ($z=-H$),
respectively. The variables $\psi_i$ stand for the streamfunctions at each depth, 
through which the horizontal flows can be expressed as
$\bm{u}_i=(u_i,v_i)=(-\partial_y \psi_i,\partial_x \psi_i)$. 

The $\delta-$PV sheets are related to the buoyancy field $b=b(x,y,z,t)$ by: 
\begin{eqnarray}
\theta_1 & = &-f\frac{b(z=0)}{N_m^2}, \\
\theta_2 & = & f \left[\frac{b(z=-h^+)}{N_m^2}-\frac{b(z=-h^-)}{N_t^2} \right],\\
\theta_3 & = & f\frac{b(z=-H)}{N_t^2}, 
\end{eqnarray}
where $f$ is the Coriolis frequency, and $N_m$ and $N_t$ are the Brunt-V\"ais\"al\"a frequencies for the mixed layer
and the thermocline, respectively.
Due to the QG assumption, the buoyancy field is also related to the streamfunction by $b=f \partial_z \psi$.
The coupling between $\theta_i$ and $\psi_i$ can be expressed, in Fourier space, by
\begin{equation}
\hat{\theta}_i =  L_{ij} \hat{\psi}_j
\label{eq:coupling}
\end{equation}
with the hat denoting the horizontal Fourier transform and $L_{ij}$ the elements of the matrix:
\begin{equation}
\displaystyle
\bm{L}=fk
\begin{pmatrix}
-\frac{\coth{\mu_m}}{N_m} & \frac{\csch{\mu_m}}{N_m} & 0 \\
\frac{\csch{\mu_m}}{N_m} & -\frac{\coth{\mu_m}}{N_m} - \frac{\coth{\mu_t}}{N_t} & \frac{\csch{\mu_t}}{N_t} \\
0 & \frac{\csch{\mu_t}}{N_t} & - \frac{\coth{\mu_t}}{N_t}
\end{pmatrix}
\label{eq:matrixL}
\end{equation}
where $k$ is the modulus of the horizontal wavenumber, and $\mu_m=N_mkh/f$ and $\mu_t=N_tk(H-h)/f$ are non-dimensional wavenumbers.

In view of the ensuing discussions, an interesting feature of this model is that it allows the computation of the horizontal 
velocity field at any depth, once the streamfunction at the discrete levels $1,2,3$ is known 
(see Appendix). 

The system is forced by mean zonal flows with $U_1=0$, $U_2=-\Lambda_m h$, $U_3=-\Lambda_m h-\Lambda_t(H-h)$ 
and mean meridional PV gradients with $\Gamma_1=f^2\Lambda_m/N_m^2$, $\Gamma_2=-f^2\Lambda_m/N_m^2+f^2\Lambda_t/N_t^2$, 
$\Gamma_3=-f^2\Lambda_t/N_t^2$. Here $\Lambda_m$ and $\Lambda_t$ account for constant vertical shear in the mixed layer 
and in the thermocline, respectively.

%%%%%%%%%%%%%%%%%%%%%%%%%%%%%%%%%%%%%%%%%%%%
\section{\label{sec:turb_flow} Turbulent flow properties}
The evolution equations (\ref{eq:model}), with (\ref{eq:coupling}) and (\ref{eq:matrixL}), are numerically 
integrated by means of a pseudospectral method on a doubly periodic square domain of side $L$ at resolution $512^2$, 
starting from an initial condition corresponding to a streamfunction whose Fourier modes have random phases and small 
amplitudes such that the resulting kinetic energy spectrum is constant in the range of wavenumbers considered.
The code was adapted from an original one developed by Ref.~\onlinecite{Smith_etal_2001} and previously used in Refs.~\onlinecite{BL_2014,FBPL2017}. 
In the model, to remove energy from the largest scales we use hypofriction with coefficient $r$, 
while small-scale dissipation (and numerical stability) is assured through an exponential filter~\cite{LaCasce1996,LaCasce1998,Smith_etal_2001}  
$D_s(\theta_i)$ acting beyond a cut-off wavenumber $k_c$.

In our numerical simulations, we adopt realistic parameter values for the (wintertime) midlatitude ocean
(similarly to Ref.~\onlinecite{CFFF2016}), as listed in table~\ref{tab:t1}. 
We further choose a domain linear size $L=500$~km and a grid spacing $\Delta x \lesssim 1$~km.
The value of the hypofriction factor is $r \simeq 2.38 \cdot 10^{-16}$~m$^{-2}$~s$^{-1}$
and the non-dimensional cut-off wavenumber for the exponential filter is $k_c=50$ (corresponding to an inverse wavelength 
of $0.1$~km$^{-1}$).
Let us mention that, even if the results presented here are in dimensional units, the numerical integration is carried out using
non-dimensional variables, in which times are made non-dimensional using the advective time-scale $L/(2\pi u)$,
where $u=0.12$~m~s$^{-1}$ is taken as the typical velocity.
%%%%%%%%%%%%%%%%%%%%%%%%%%%%%
% Tab. model parameters
\begin{table}
 \caption{\label{tab:t1} Main physical parameters of the model; in the present study $\Lambda_m=\Lambda_t=\Lambda$.}
 \begin{ruledtabular}
  \begin{tabular}{ccc}
    Vertical shear & $\Lambda$ & $10^{-4}$~s$^{-1}$\\
    Mixed-layer buoyancy frequency & $N_m$ & $2 \cdot 10^{-3}$~s$^{-1}$\\
    Thermocline buoyancy frequency & $N_t$ & $8 \cdot 10^{-3}$~s$^{-1}$\\
    Coriolis frequency &  $f$ & $10^{-4}$~s$^{-1}$\\
\end{tabular}
\end{ruledtabular}
\end{table}
%%%%%%%%%%%%%%%%%%%%%%%%%%%%%
% Tab. models TC, ML, F
\begin{table}
 \caption{\label{tab:t2} Depth values used for the three different models (see text).}
 \begin{ruledtabular}
  \begin{tabular}{cccc}
    Model & TC & ML & F\\
    \hline
    Mixed-layer depth $h$ & 250~m & 100~m & 100~m\\
    Total depth  $H$ &  500~m & 1000~m & 500~m\\
\end{tabular}
\end{ruledtabular}
\end{table}
%%%%%%%%%%%%%%%%%%%%%%%%%%%%%

We consider three cases: the thermocline only case (TC), the mixed-layer only case (ML), the full model (F), which  
are specified by the values of the depths $h$ and $H$ in table~\ref{tab:t2}. For the thermocline only case, 
the parameter $h$ does not have a physical meaning and its value is set to $H/2$; moreover we take $N_m=N_t=8 \cdot 10^{-3}$~s$^{-1}$. 
In the following we will focus on dynamics down to depth $z=-500$~m, for all the three considered cases.

%%%%%%%%%%%%%%%%%%%%%%%%%%%%%%%%%%%%%%%%%%%%
\subsection{\label{sec:ke_turb} Spatial structure and kinetic energy spectra}
\begin{figure*}
  \includegraphics[scale=0.85]{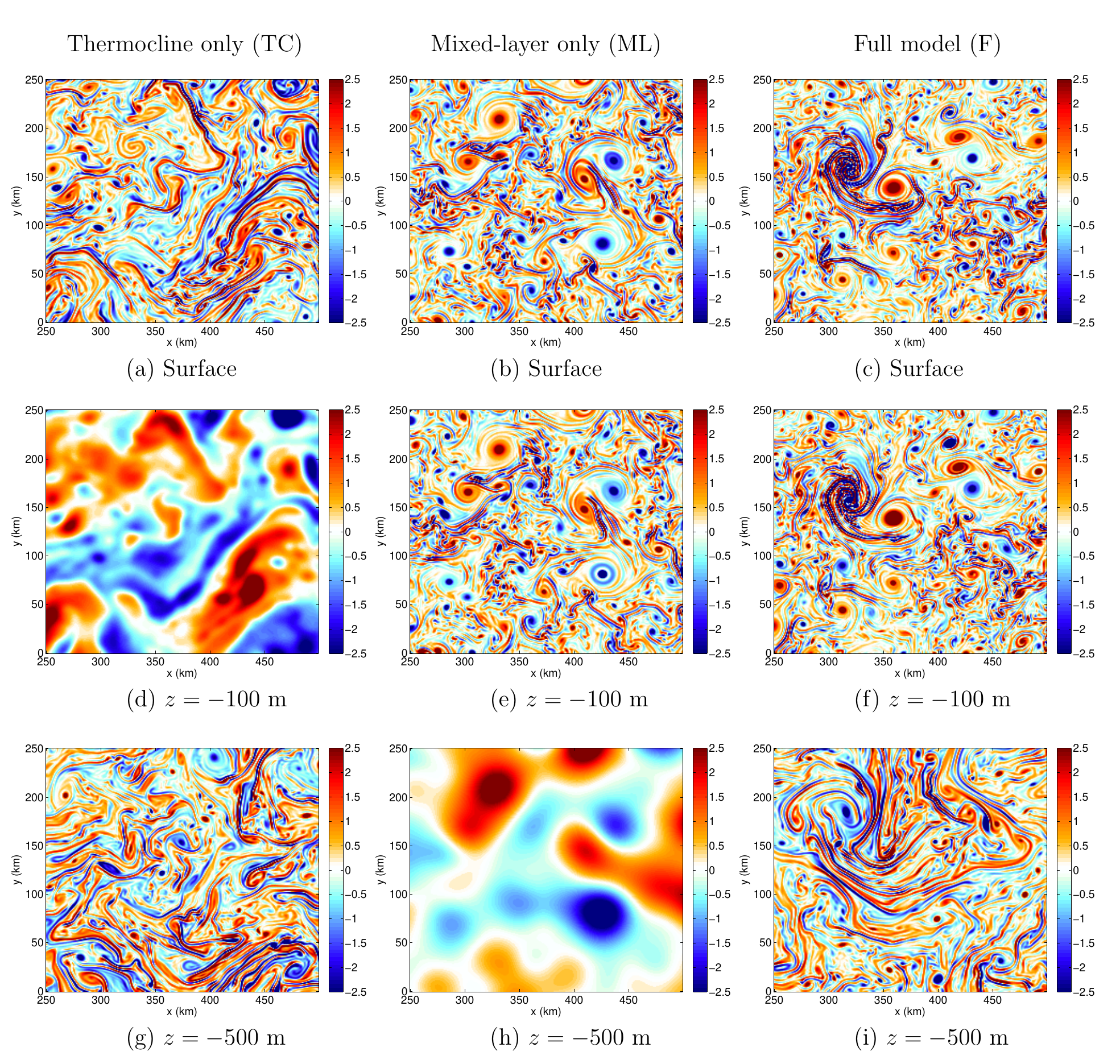}
  \caption{\label{fig:f2} Vorticity field, normalized by its rms value, for the TC, ML and F cases (columns from left to right)
  %at $z=0,-100,-500$~m (from top to bottom). 
  at the surface (a, b, c), at $z=-100$~m (d, e, f) and at $z=-500$~m (g, h, i). 
  In all panels, a zoom on the subregion $[250,500] \times [0,250]$~km$^2$ 
  of the total domain is shown.}
\end{figure*}

The spatial organization of the horizontal flow can be inspected by plotting the vorticity field 
$\zeta=\partial_x v - \partial_y u=(\partial_x^2 + \partial_y^2) \psi$. 
Some snapshots of $\zeta$ at a given time after the system reached the statistically steady state 
are shown in Fig.~\ref{fig:f2}, after normalization by the root-mean-square (rms) value 
$\zeta_\mathrm{rms}=\langle \zeta^2 \rangle^{1/2}$ (with brackets indicating a spatial average). 
In the figure, each column corresponds to a different model (TC, ML, F, from left to right). 
In the top, middle and bottom row $\zeta/\zeta_\mathrm{rms}$ is shown at the surface,
at $z=-100$~m and at $z=-500$~m, respectively.

In all the examined cases, the surface fields are characterized by a whole range of active scales.
In the TC case, vorticity is prominently organized in a tangle of long filaments (Fig.~\ref{fig:f2}a). 
Eddies of different sizes are also present, so that the flow field is characterized by both features simulatenously. 
As argued in Ref.~\onlinecite{CFFF2016}, in this case the dynamics of surface buoyancy anomalies decouple from those at the bottom
at sufficiently small scales, essentially giving rise to turbulent flows of surface quasi-geostrophic (SQG) type.
Then, as only the mesoscale instability is here present, small-scale eddies are generated by a roll-up instability
of larger flow features~\cite{Held_etal_1995}.
Small-scale eddies rapidly decay with depth and at $z=-100$~m only the largest structures are still present (Fig.~\ref{fig:f2}d).
At $z=-500$~m the TC vorticity field is statistically equivalent
to its surface counterpart, due to SQG-like dynamics at level $z=-H$ and symmetry of the dynamics with respect to the
half total depth.
%%%%%%%%%%%%%%%%%%%%%%%%%%%%
% Fig. vorticity snapshots (zoom) 
% z=0, -h, -H
% TC, ML, F 
%%%%%%%%%%%%%%%%%%%%%%%%%%%%

The situation is quite different for the ML case (central column of Fig.~\ref{fig:f2}).
In the presence of mixed-layer instabilities, eddies, initially of the same size as the scale of the submesoscale
instability, grow until being balanced, in a statistical sense, by hypofriction. Several coherent vortices
are visible in the surface vorticity field.
One can also remark that filaments are now shorter and less intense. 
In sharp contrast with the TC case, vorticity varies very little, in a statistical sense, over the first $100$ meters 
below the surface, pointing to energetic submesoscales in this whole depth range.
Below the mixed layer, small scales decay in a way similar to what observed for TC,
until at $z=-500$~m only very large vorticity patches are found. Notice that here the thermocline
is virtually absent, as $H=10h$, which accounts for the differences observed at $z=-500$~m with respect to other cases.

The picture in the full model (right column of Fig.~\ref{fig:f2}) is similar to that of the ML case in the
mixed layer and the upper thermocline. However, at larger depths, the flow recovers energy at small scales,
due to the effect of the finite-depth thermocline and SQG-like dynamics at its bottom.
At $z=-500$~m, vorticity has a rather filamentary structure in which several small eddies are immersed.
%%%%%%%%%%%%%%%%%%%%%%%%%%%%
% Fig. kinetic energy spectra 
% versus depth
% TC, ML, F 
\begin{figure*}
  \includegraphics[width=\textwidth]{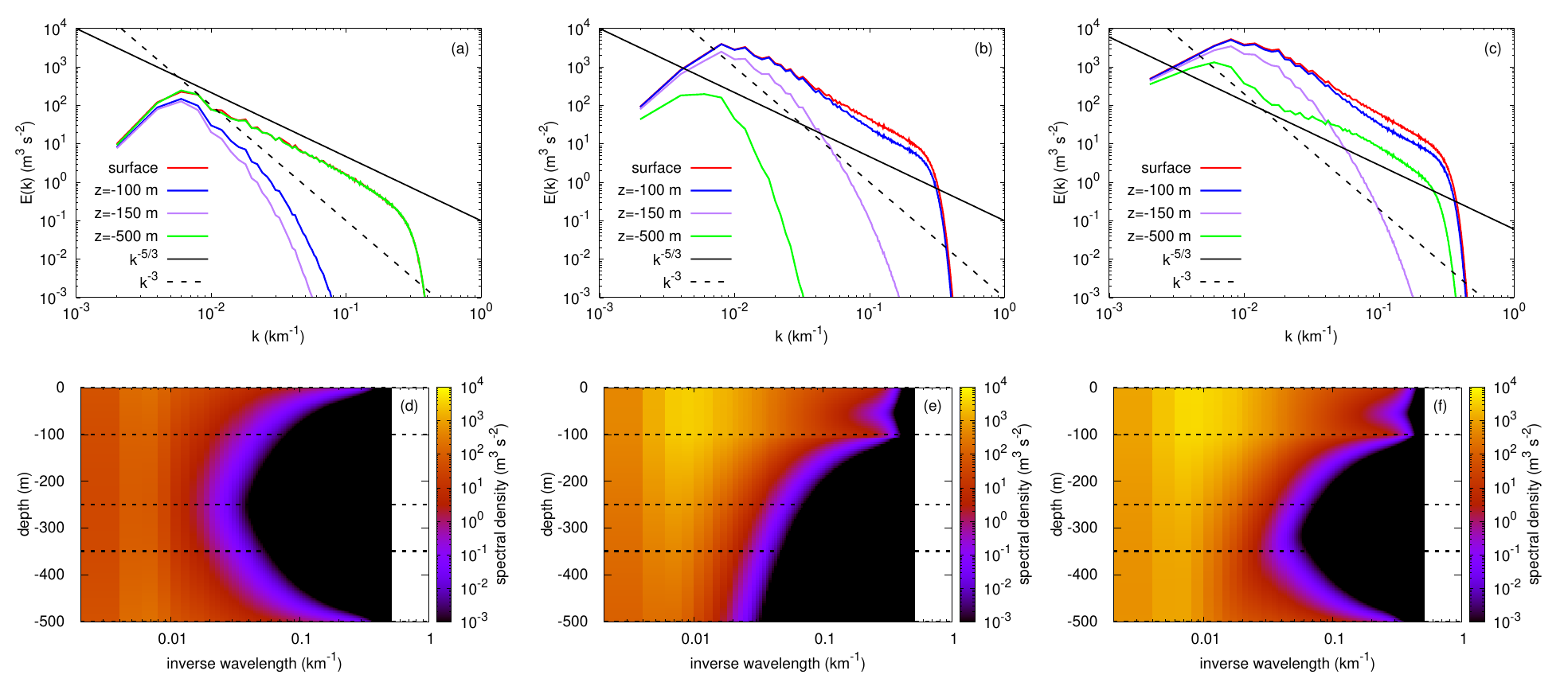}
  \caption{\label{fig:f3} Kinetic energy spectra, temporally averaged over several flow realizations in the 
  statistically steady state, at some selected depths for the TC (a), ML (b) and F (c) cases. 
  The variation of spectra with depth is shown in (d), (e), (f) for the TC, ML and F cases, respectively. 
  Here the dashed lines indicate the reference depths considered for the horizontal Lagrangian dispersion (Sec.~\ref{sec:disp_h}).} 
\end{figure*}
%%%%%%%%%%%%%%%%%%%%%%%%%%%%

A statistical characterization of these turbulent flows can be provided by their kinetic energy spectrum $E(k)$, which is shown 
in Fig.~\ref{fig:f3} for each model.
The top row shows the spectra (as a function of the horizontal wavenumber)
at some selected depths, while in the bottom row a more complete description for varying depth is reported. 
Here the dashed lines correspond to the reference depths considered for the horizontal Lagrangian dispersion (Sec.~\ref{sec:disp_h}).  

In the TC case the spectrum at the surface (and at the bottom) displays a scaling behavior that is rather close to $E(k) \sim k^{-5/3}$,
as expected in SQG turbulence.
In the interior, the energetic content
of the largest scales is comparable to the corresponding value at the surface, but the spectrum rapidly falls off,
due to the decay of small eddies with depth. Indeed, already at $z=-100$~m, $E(k)$ is found to be definitely
steeper than $k^{-3}$.

In the ML case, kinetic energy spectra are similar in a broad range of wavenumbers over the mixed layer, 
which is then fully energized. At the surface and at the base of the mixed layer 
their scaling is not far from $k^{-5/3}$,
though slightly steeper at large scale, as also observed in Ref.~\onlinecite{CFFF2016}.
Below $z=-100$~m, the spectrum shows a fast decrease with the wavenumber and becomes steeper than $k^{-3}$,
due to less and less intense small scales at larger and larger depths. 
We remark that we verified that the value of the total depth $H$ 
does not considerably affect the spectral properties of the turbulent flows down to $z=-500$~m. 

When both the mixed layer and the thermocline are present (case F), $E(k)$ is very similar to the spectrum found
in the previous case, both in the mixed layer and in the upper thermocline. Nevertheless, close to $z=-500$~m
it displays energetic small scales again, due to the dynamics at the bottom.
At this depth, similarly to what occurs at the surface, a scaling range with spectrum not far from $k^{-5/3}$ is observed
at relatively small scales, while at large scales the spectrum is steeper and tends to approach $k^{-3}$. 

%%%%%%%%%%%%%%%%%%%%%%%%%%%%%%%%%%%%%%%%%%%%
\subsection{\label{sec:turb_int} Turbulence intensity at varying depth}
Here we consider the variation with depth of the typical intensities of the turbulent flow velocities and of their gradients, 
both of which are expected to be relevant for the transport of Lagrangian particles. 

\begin{figure}
%  \centering
  \includegraphics[width=0.9\columnwidth]{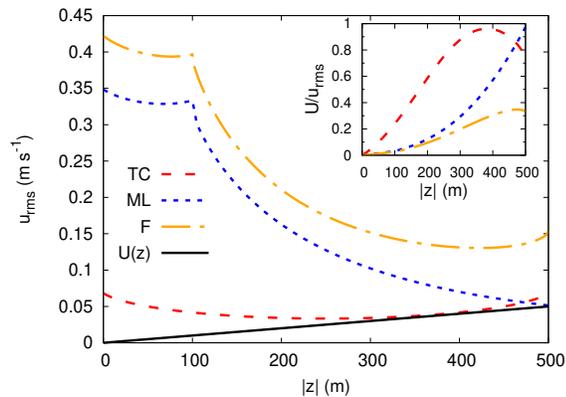}
  \caption{\label{fig:f4} Typical turbulent velocity fluctuations $u_\mathrm{rms}$ 
   and mean flow intensity $U(z)$ 
   as a function of depth for the 
   TC, ML and F cases. 
   Inset: ratio $U(z)/u_\mathrm{rms}$ versus depth. The rms velocities shown here
   are temporally averaged over several flow realizations in the statistically steady state.}
\end{figure}

We first examine the rms turbulent velocity $u_\mathrm{rms}=\langle |\bm{u}|^2 \rangle^{1/2}$, and compare it 
to the intensity of the zonal mean flow $\bm{U}=U(z)\hat{\bm{x}}=\Lambda z\hat{\bm{x}}$ (see Sec.~\ref{sec:model} 
and table~\ref{tab:t1}). 
The behavior of both quantities as a function of the depth $|z|$ is reported in Fig.~\ref{fig:f4},   
where the inset shows the (inverse) turbulence intensity, $U/u_\mathrm{rms}$, versus depth. 
Generally speaking, one can see from these plots that turbulence becomes weaker, while the mean flow gains importance, 
with depth.
The way this occurs, however, depends on the model dynamics. While for the TC and ML cases the mean flow can become comparable
to typical turbulent velocity fluctuations, in the full model $U/u_\mathrm{rms}$ never exceeds $0.4$.
In the absence of the mixed layer, $U>u_\mathrm{rms}/2$ at depths larger than $250$~m 
(half the total depth, where the turbulent flow is weakest), and the ratio $U/u_\mathrm{rms}$ reaches its peak 
value ($\approx 1$) close to the bottom, namely for $300$~m$<z<400$~m, then slightly decreasing to reach $0.7$ 
at $z=-500$~m, due to more intense turbulence at the bottom. 
In the presence of the mixed layer, $u_\mathrm{rms}$ remains essentially 
unchanged in the first $100$ meters from the surface. Below the mixed layer it decreases, but the mean flow becomes 
comparatively relevant only at quite large depths (in the ML case $U>u_\mathrm{rms}/2$ at $|z|>400$~m).
Moreover, while in the ML case $U/u_\mathrm{rms} \simeq 1$ at $z=-500$~m, in the full model the more energetic
turbulent dynamics at the bottom of the thermocline partially compensate the importance of the mean flow at the
largest depths, where $U/u_\mathrm{rms} \approx 0.35$ at most.
%%%%%%%%%%%%%%%%%%%%%%%%%%%%
% Fig. urms 
% versus depth
% TC, ML, F 
%%%%%%%%%%%%%%%%%%%%%%%%%%%%

\begin{figure}
%  \centering
  \includegraphics[width=0.9\columnwidth]{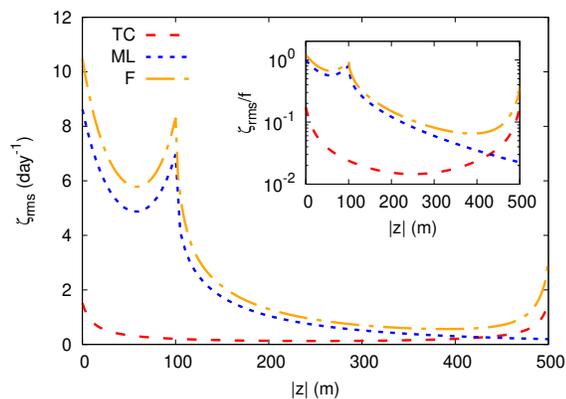}
  \caption{\label{fig:f5} Typical turbulent vorticity fluctuations $\zeta_\mathrm{rms}$ as a function of depth 
   for the TC, ML and F cases. 
   In the inset $\zeta_\mathrm{rms}$ is normalized by the Coriolis frequency $f$ and plotted in semilogarithmic scale. 
   The rms vorticities shown here are temporally averaged over several flow realizations in the statistically steady state.}
\end{figure}

If the rms velocity gives information about the intensity of the turbulent flow, what matters in the separation process
of advected Lagrangian particles are the velocity gradients. 
The latter can be quantified by the rms vorticity $\zeta_\mathrm{rms}(z)$,
which is shown in Fig.~\ref{fig:f5} for the three models. Its decrease with depth is evident in all cases.
In the TC case, as for $u_\mathrm{rms}$, the symmetric behavior with respect to $|z|=250$~m results from the dynamics at the bottom.
The trace of the latter is also visible in the full model, where it causes an increase of $\zeta_\mathrm{rms}$
at the largest depths.
This feature is absent in the ML case, where $\zeta_\mathrm{rms}$ monotonously decreases below the mixed layer.
In the presence of the latter (ML and F cases), the rms vorticity is always larger above $z=-100$~m than
deeper below, and typically larger than in the TC case. 
%%%%%%%%%%%%%%%%%%%%%%%%%%%%
% Fig. zrms 
% versus depth
% TC, ML, F 
%%%%%%%%%%%%%%%%%%%%%%%%%%%%

%%%%%%%%%%%%%%%%%%%%%%%%%%%%%%%%%%%%%%%%%%%%
\section{\label{sec:disp} Lagrangian pair dispersion}
In the following we will consider the horizontal dispersion properties of an ensemble of Lagrangian
tracer particles moving at fixed depth $z^*$ in the turbulent flows produced by the TC, ML and F models.
The equation of motion of these particles is
\begin{equation}
\frac{d{\bm x}_i}{dt}=\bm{v}({\bm x}_i(t),z^*,t), \quad \; i=1,...,N,
\label{eq:lagr_motion}
\end{equation}
where ${\bm x}_i=(x_i,y_i)$ denotes the horizontal position of particle $i$ and $\bm{v}(x,y,z^*,t)=\bm{u}(x,y,z^*,t)+\bm{U}(z^*)$
the total velocity field at the particle position (at depth $z^*$) resulting from the sum of the turbulent component
$\bm{u}$, computed from the streamfunction in Eqs.~(\ref{eq:psi_z1}-\ref{eq:psi_z2}), 
and the mean flow $\Lambda z^* \hat{\bm{x}}$. 

In our numerical experiments, Eq.~(\ref{eq:lagr_motion}) is integrated using a fourth-order Runge-Kutta scheme 
and bicubic interpolation in space of the velocity field at particle positions~\cite{Hua1994}. 
We assume that the particle motion occurs in an infinite domain and use the spatial periodicity of the Eulerian flow  
to compute the Lagrangian velocities outside the computational box.

The particles are seeded in the 
turbulent flows once the latter have reached statistically steady conditions.
At each considered depth particles are initially placed in triplets,
uniformly spread (on the horizontal) over the spatial domain. The number of triplets is $M=64 \times 64=4096$
at each depth level. Each triplet is constituted by a pair along $x$ and one along $y$, both of which are
characterized by an initial separation $R_0=\Delta x/2 \simeq 500$~m (with $\Delta x \simeq 1$~km the grid spacing). 
For simplicity, below we will focus on the results
from the indicators based on the total separation $R=\sqrt{R_x^2+R_y^2}$ (where $R_x$ and $R_y$ are
the separations along $x$ and $y$, respectively), for pairs initially along $x$. 
We verified that there was no major difference in the statistics when
considering dispersion in the $x$ or $y$ direction,
despite the presence of the mean zonal shear at depth.
In this study we only consider original pairs, and we choose as reference depths $z=0,-100,-250,-350,-500$~m, 
except where explicitly mentioned.

In Sec.~\ref{sec:disp_h} we examine horizontal dispersion at different depths using  
both fixed-time indicators, as relative dispersion (as a function of time)~\cite{ABCCV1997,LaCasce2008,FBPL2017} 
and fixed-scale ones, as the finite-size Lyapunov exponent (FSLE, or FSLE-I)~\cite{ABCCV1997,ABCPV1997,CV2013}. 
In Sec.~\ref{sec:fsle-ii}, we address the 
properties of the relative motion of subsurface particles
with respect to surface ones,
by analyzing the so-called FSLE of the $2^\mathrm{nd}$ kind (FSLE-II)~\cite{ILRSV2002,LCFS2019}.  

%%%%%%%%%%%%%%%%%%%%%%%%%%%%%%%%%%%%%%%%%%%%
\subsection{\label{sec:disp_h} Horizontal dispersion}
Here we are interested in assessing how the horizontal dispersion process varies in the vertical. In particular we aim at 
identifying different dynamical regimes and at higlighting possible transitions among them as a function of depth. 

The first diagnostic we consider is relative dispersion, which is defined as
\begin{equation}
\langle R^2(t) \rangle = \langle |\bm{x}_i(t) - \bm{x}_j(t)|^2 \rangle,
\label{eq:reldisp}
\end{equation}
where the average is over all pairs $(i,j)$ such that at $t=0$ (the release time) $|\bm{x}_i(0) - \bm{x}_j(0)| = R_0$.
%%%%%%%%%%%%%%%%%%%%%%%%%%%%
% Fig. reldisp
% normalized (t/zrms)
% at some depths: 
% z=0,-100,-250, -350, -500 (in m)
% TC, ML, F 
\begin{figure*}
%  \centering
  \includegraphics[width=0.33\textwidth]{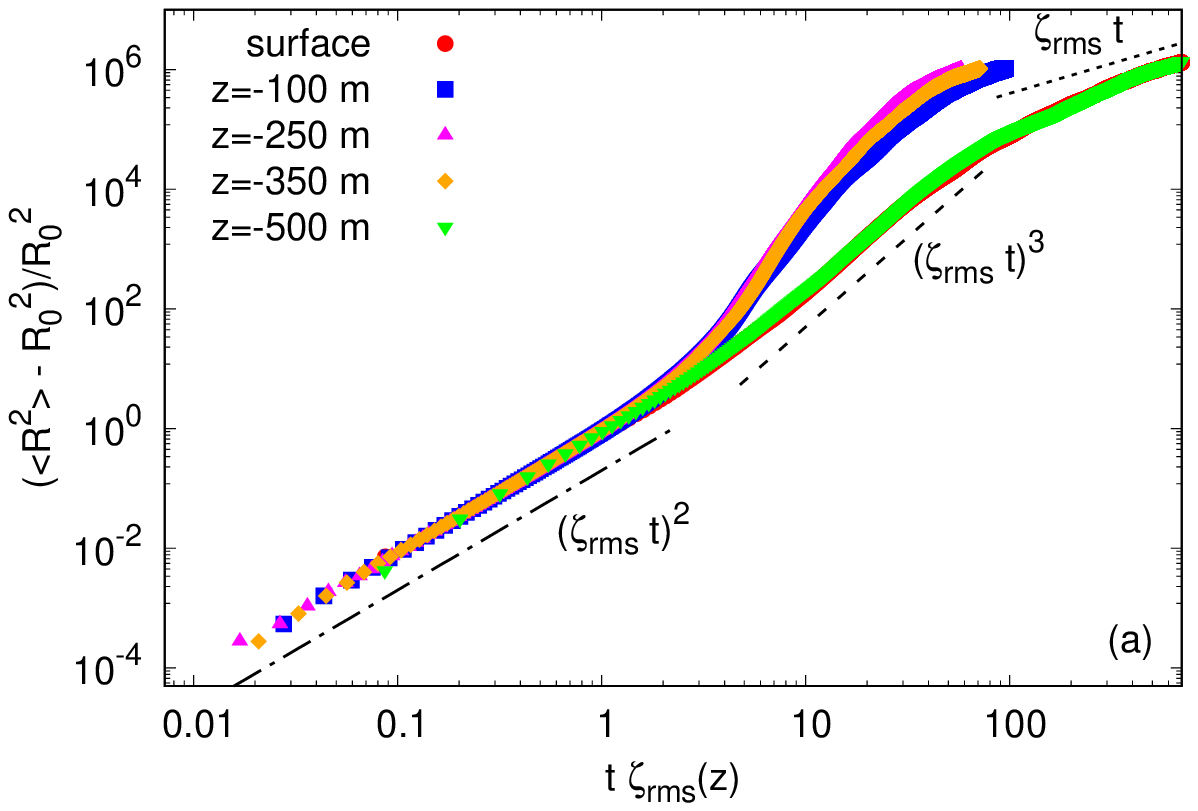}
  \includegraphics[width=0.33\textwidth]{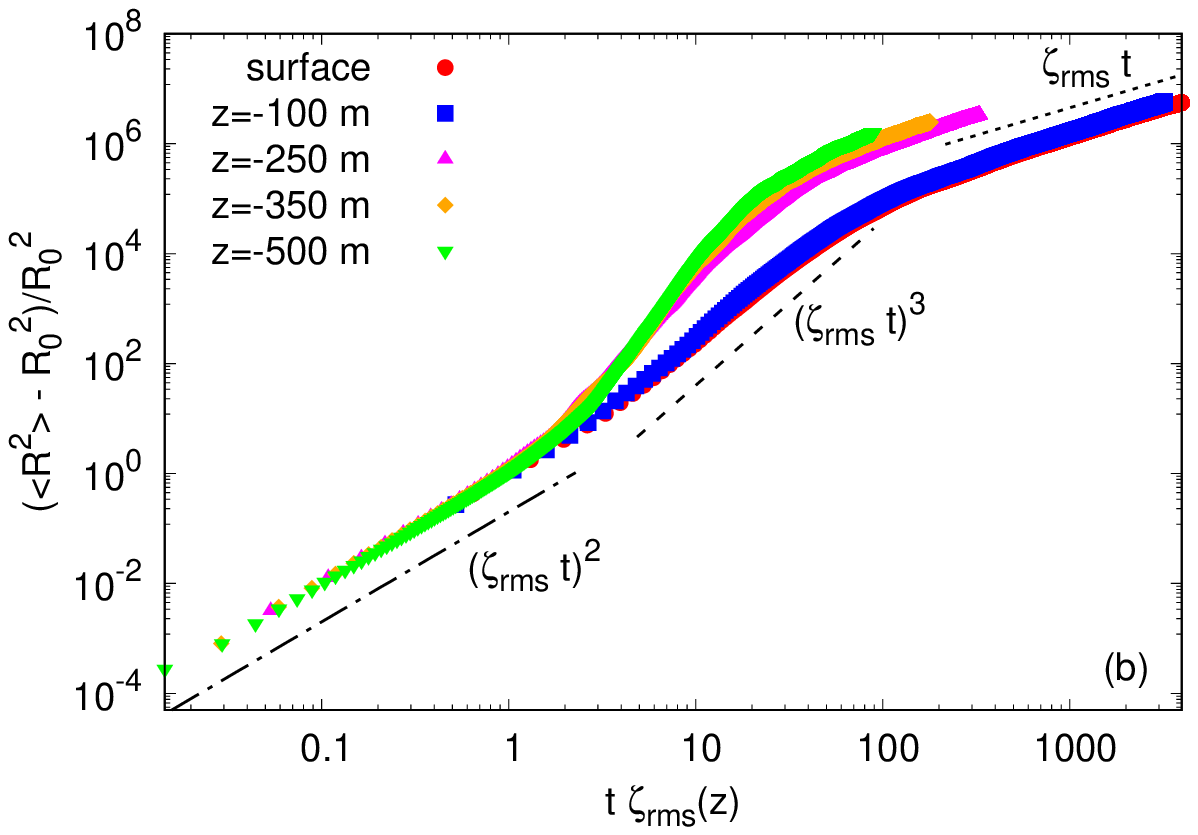}
  \includegraphics[width=0.33\textwidth]{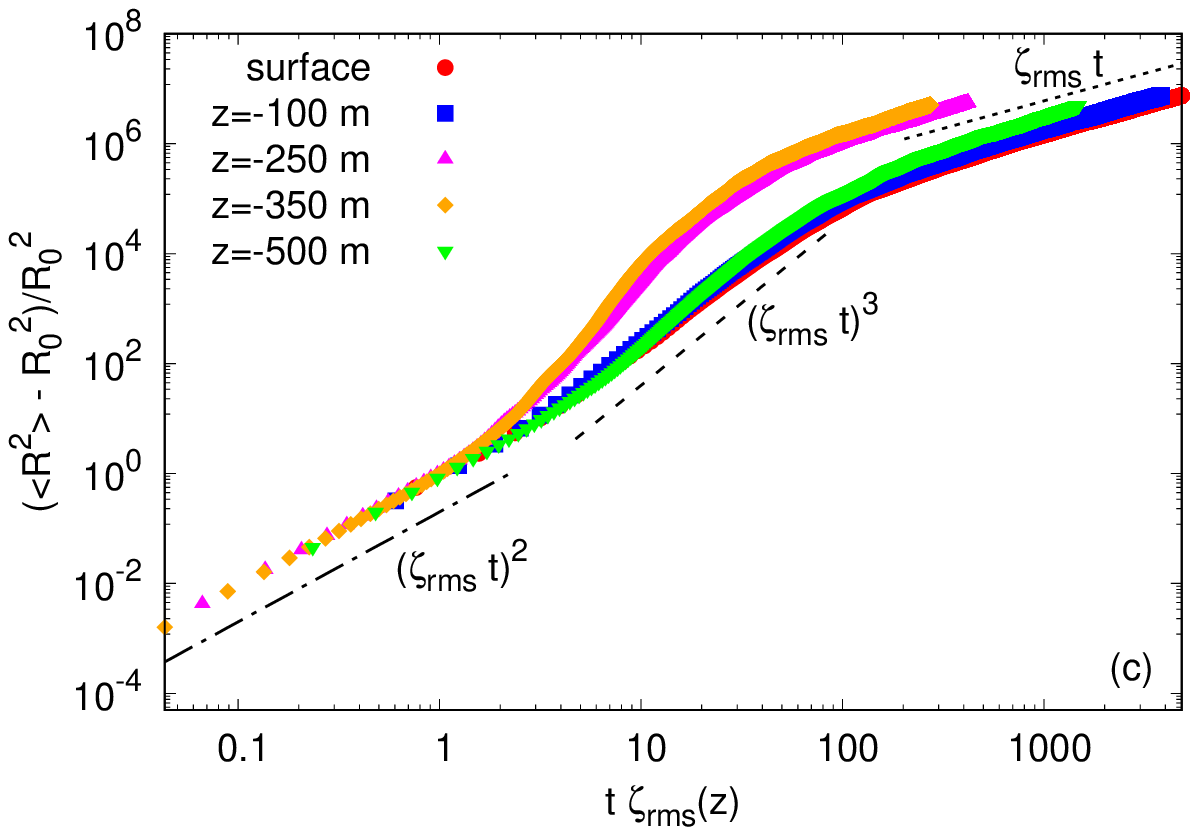}
  \caption{\label{fig:f6} Relative dispersion (after subtraction of the initial value $R_0^2$ and normalization by it)
  as a function of the time rescaled by the rms vorticity $\zeta_\mathrm{rms}$ (Fig.~\ref{fig:f5}) 
  at the reference depths 
  for the TC (a), ML (b) and F (c) cases.
  }  
\end{figure*}
%%%%%%%%%%%%%%%%%%%%%%%%%%%%

Assuming that relative velocity is independent of the particle pair separation, at sufficiently short times 
a ballistic behavior $\langle R^2 \rangle - R_0^2 \sim t^2$ is 
expected~\cite{Batchelor1950,BBRS1990,BOXBB2006,FBPL2017,Spydell_etal_2020}.
At intermediate times, for which dispersion scales are within the inertial range of the turbulent cascade,
the expected behavior depends on the form of the kinetic energy spectrum 
(see Ref.~\onlinecite{Bourgoin2018} for a compact review).  
Assuming a power-law spectrum, $E(k) \sim k^{-\beta}$,
the value of the exponent $\beta$ then determines the expected dispersion regime. For a rough flow, for which $\beta<3$,
relative dispersion should scale as $\langle R^2 \rangle \sim t^{4/(3-\beta)}$ (see Refs.\onlinecite{BCCV1999,BBCDLT2005,LaCasce2008}),
which includes Richardson superdiffusive behavior $\langle R^2(t) \rangle \sim t^3$ for $\beta=5/3$.
In such a case ($\beta<3$) the dispersion process is referred to as a local one, meaning that the growth
of the separation distance between two particles in a pair is governed by eddies of the same size as the separation
itself~\cite{LaCasce2008,FBPL2017}. When $\beta>3$, instead, the flow is smooth and the expectation for
relative dispersion is $\langle R^2(t) \rangle \sim \exp{(2 \, \lambda_L \, t)}$ (see Refs.~\onlinecite{FGV2001,LaCasce2008}),
where $\lambda_L$ is the Lagrangian maximum Lyapunov exponent. Such an exponential growth of $\langle R^2 \rangle$
is typically referred to as a nonlocal dispersion regime, meaning governed by the largest eddies~\cite{LaCasce2008,FBPL2017}.
Finally, for separations much larger than the largest characteristic flow scale, a diffusive behavior
$\langle R^2 \rangle \sim t$ is expected, due to essentially uncorrelated particle velocities.

To identify different dispersion regimes, it is useful to perform a rescaling of the considered variables. 
A relevant quantity is, in this respect, the rms vorticity $\zeta_\mathrm{rms}$, which accounts for the intensity 
of typical velocity gradients. 
The behavior of $\langle R^2(t) \rangle$ is reported in Fig.~\ref{fig:f6}. Here time is rescaled with 
$1/\zeta_\mathrm{rms}$, which provides an estimate of the typical time over which trajectory pairs loose memory of their 
initial condition; relative dispersion is plotted after subtraction of its initial value $R_0^2$ and normalization by 
the latter.
Through this representation we are able to detect several distinct behaviors, which correspond to the different dispersion 
regimes that are realized in the course of time. The quite nice collapse of the data further indicates the generality of 
the observed spreading mechanisms.
As it can be seen, independently of the model and of the depths, 
when $t$ is smaller than a time of order $1/\zeta_\mathrm{rms}$, 
a clear ballistic behavior ($\sim t^2$) is found.
In the opposite limit of very large times, all curves indicate diffusive behavior ($\sim t$), as expected.
At intermediate times, relative dispersion approaches a $t^3$ scaling, suggesting Richardson local dispersion, 
at the surface (for all models), at the base of the mixed layer (for the ML and F cases) and at the bottom of the 
thermocline (for the TC and F models). These results are in fair agreement with the expectation based on the shape of the 
kinetic energy spectrum, which in these cases is close to $E(k) \sim k^{-5/3}$ (Fig.~\ref{fig:f3}). 
For the remaining cases (\ie~in the interior of the TC system, below the mixed layer in the ML one, and in the upper thermocline 
for the F case), the collapse of the curves (for fixed model and different depths), points to a common dispersion regime 
characterized by fast growth in time (meaning faster than $t^3$) of $\langle R^2 \rangle$. 
Here, based on $E(k)$ being steeper than $k^{-3}$, we should expect exponential growth of the squared separation 
distance (nonlocal dispersion). 
Even when examined on a lin/log scale, the data, however, do not quantitatively support this picture 
and do not allow to measure $\lambda_L$ (not shown). 
A possible reason for such a difficulty is that relative dispersion is constructed as an average at fixed time~\cite{CV2013}.
Indeed, in the presence of large variability as a function of the initial pair location and/or time, as it is found to be 
the case here (not shown), $\langle R^2(t) \rangle$ does not allow the detection of the correct scaling behavior. 
An illustration of this effect showing a spurious anomalous regime for a system of point vortices is documented 
in Ref.~\onlinecite{BCCLV2000}, while Ref.~\onlinecite{BBCDLT2005} reports the difficulty to detect Richardson’s scaling 
from $\langle R^2(t) \rangle$, but not from the FSLE, in direct numerical simulations
of three-dimensional homogeneous isotropic turbulence.
%%%%%%%%%%%%%%%%%%%%%%%%%%%%
% Fig. reldiff (rescaled) 
% at some depths: 
% z=0,-100,-250, -350, -500 (in m)
% TC, ML, F 
\begin{figure*}
%  \centering
  \includegraphics[width=0.33\textwidth]{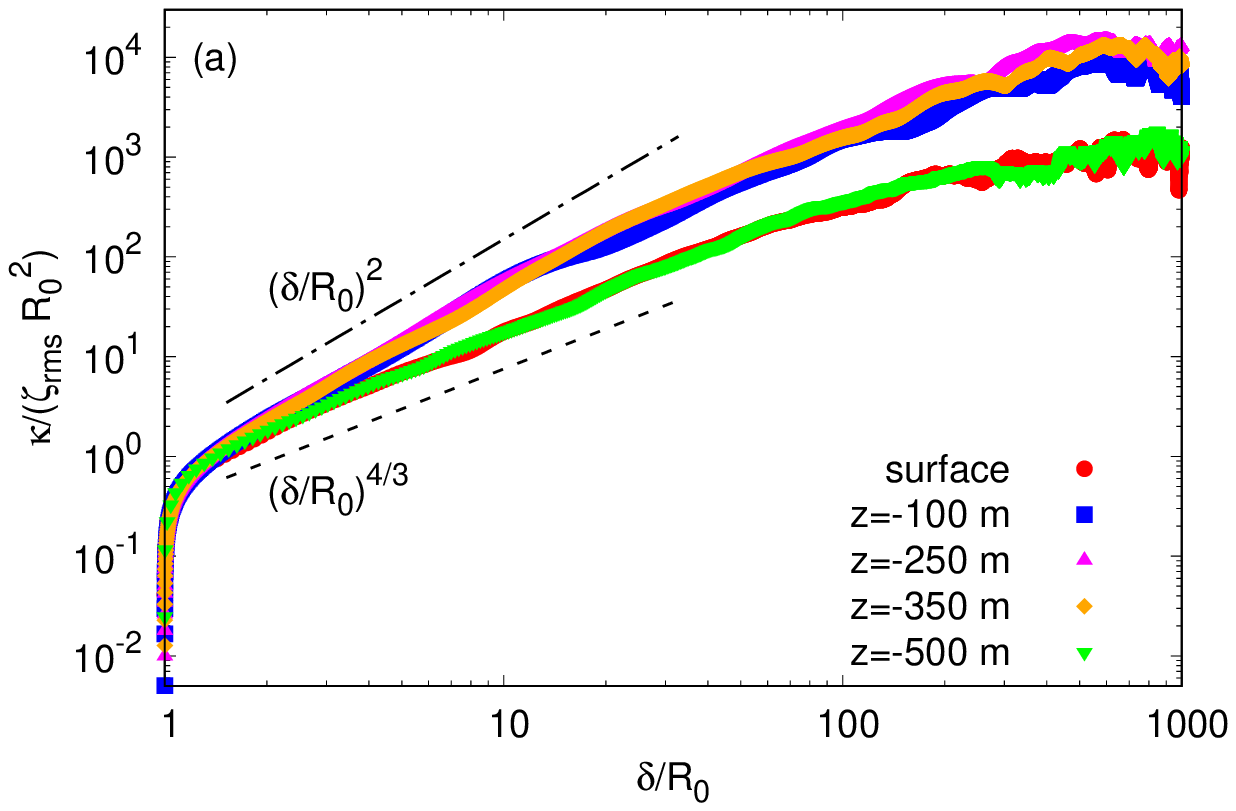}
  \includegraphics[width=0.33\textwidth]{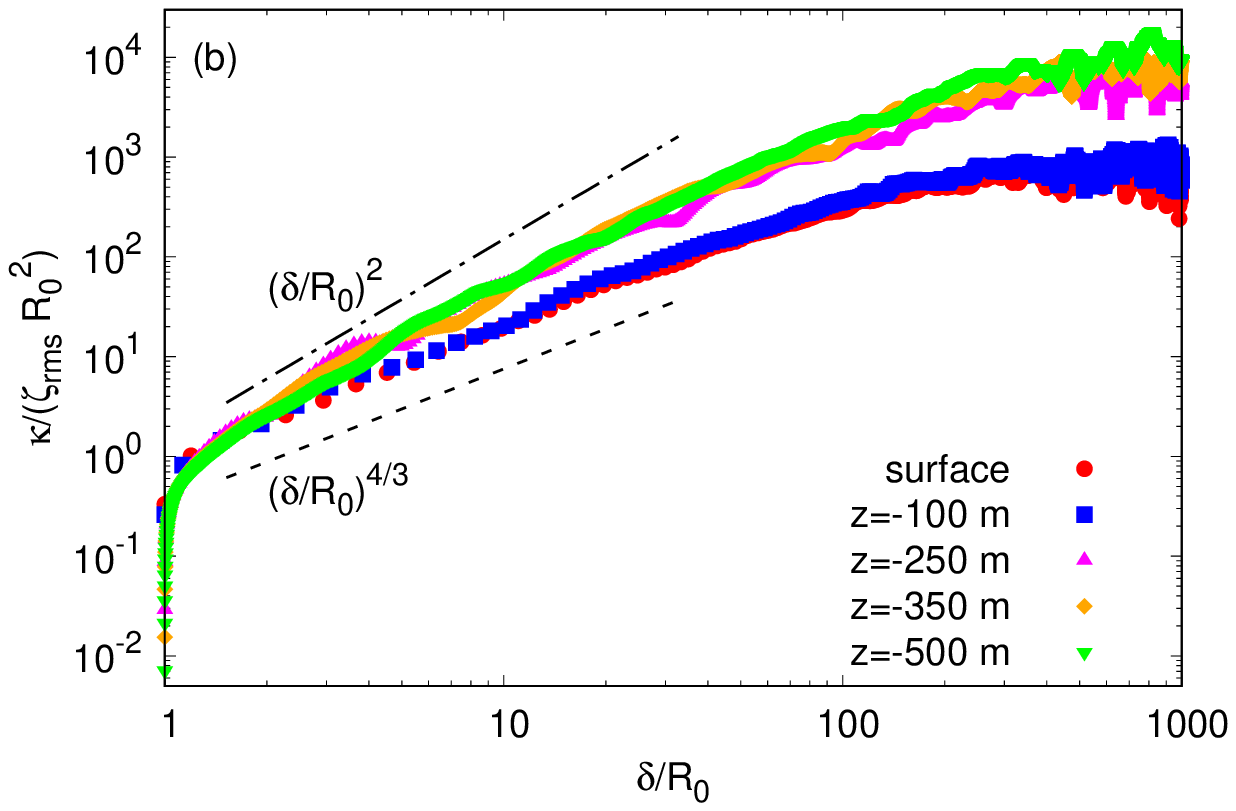}
  \includegraphics[width=0.33\textwidth]{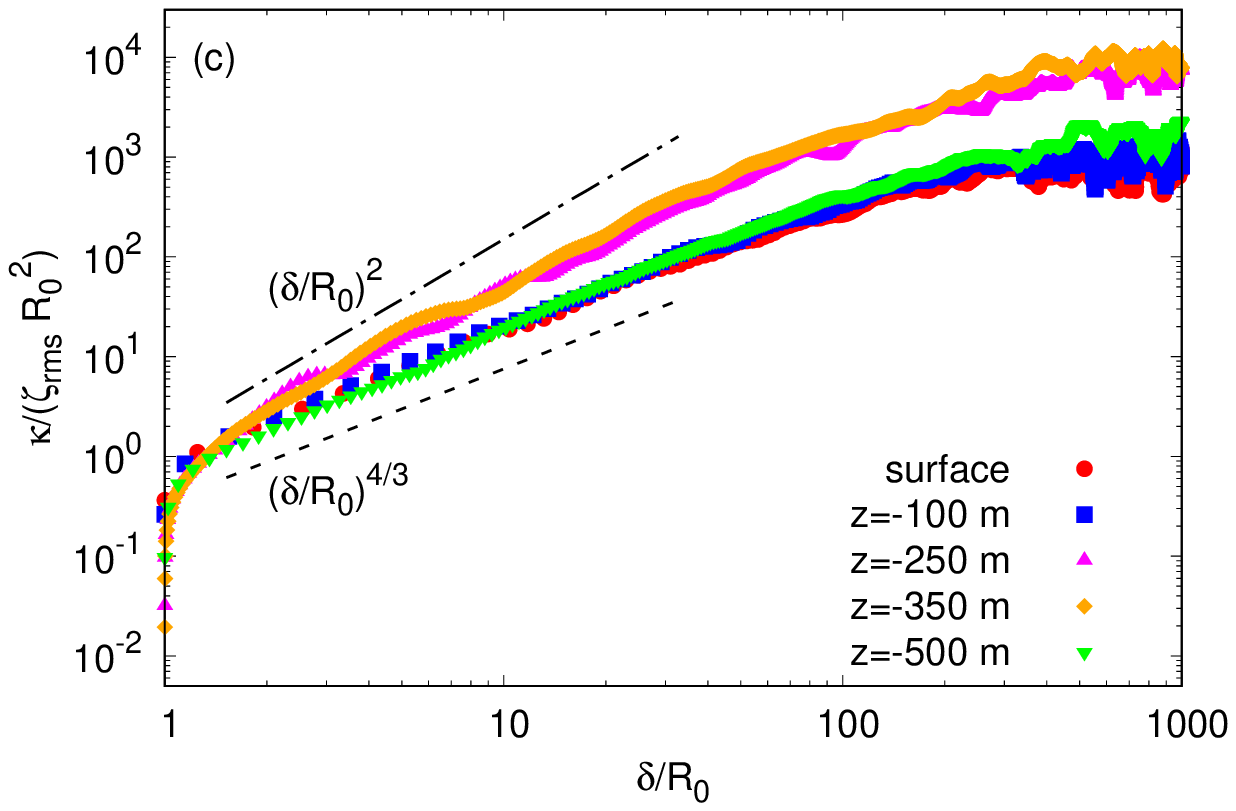}
  \caption{\label{fig:f7} Relative diffusivity as a function of the separation distance $\delta=\langle R^2(t) \rangle^{1/2}$, 
  after rescaling time with $\zeta_\mathrm{rms}^{-1}$ and distance with $R_0$, 
  at the reference depths for the TC (a), ML (b) and F (c) cases.
  The dashed and dash-dotted lines respectively correspond to $\delta^{4/3}$ (Richardson local regime) 
  and $\delta^2$ (nonlocal regime) for reference.}
\end{figure*}
%%%%%%%%%%%%%%%%%%%%%%%%%%%%
%%%%%%%%%%%%%%%%%%%%%%%%%%%%
% Fig. FLSE-I 
% normalized by zrms 
% at some depths: 
% z=0,-100,-250, -350, -500 (in m)
% TC, ML, F 
\begin{figure*}
%  \centering
  \includegraphics[width=0.33\textwidth]{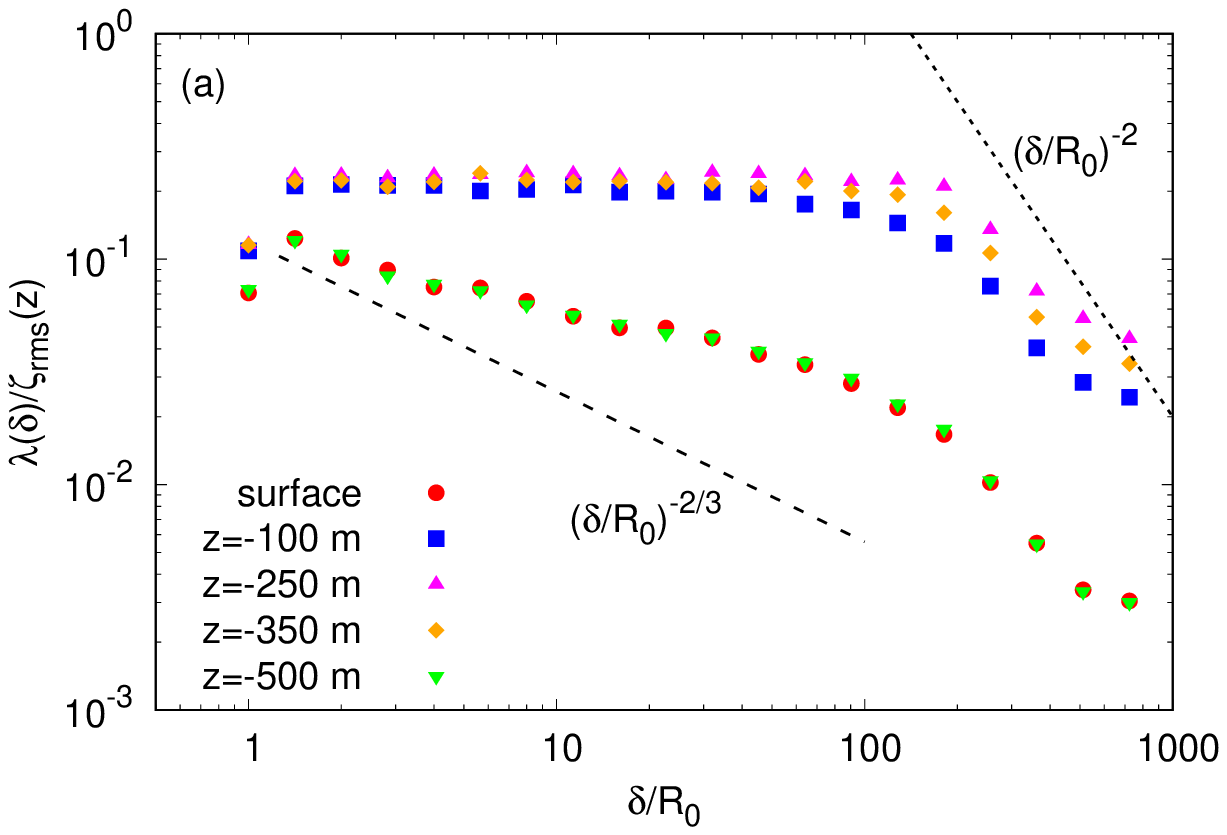}
  \includegraphics[width=0.33\textwidth]{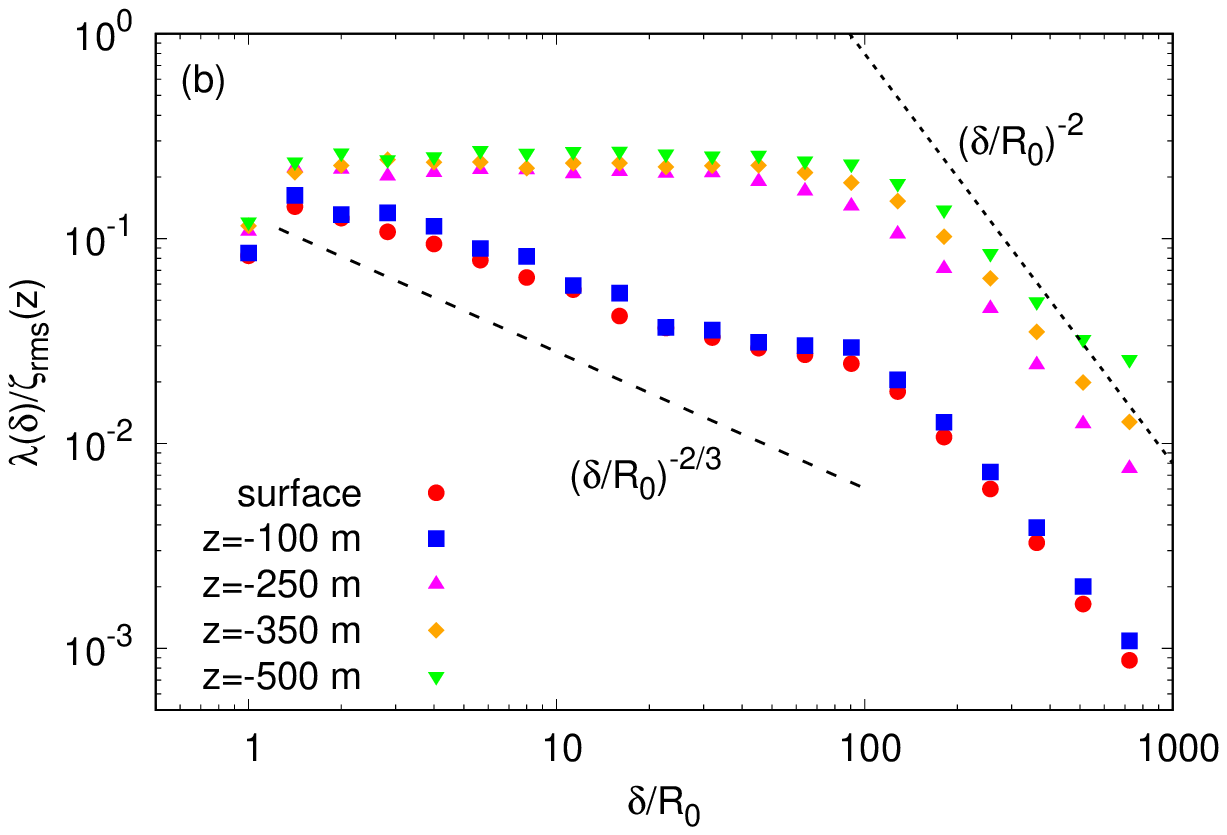}
  \includegraphics[width=0.33\textwidth]{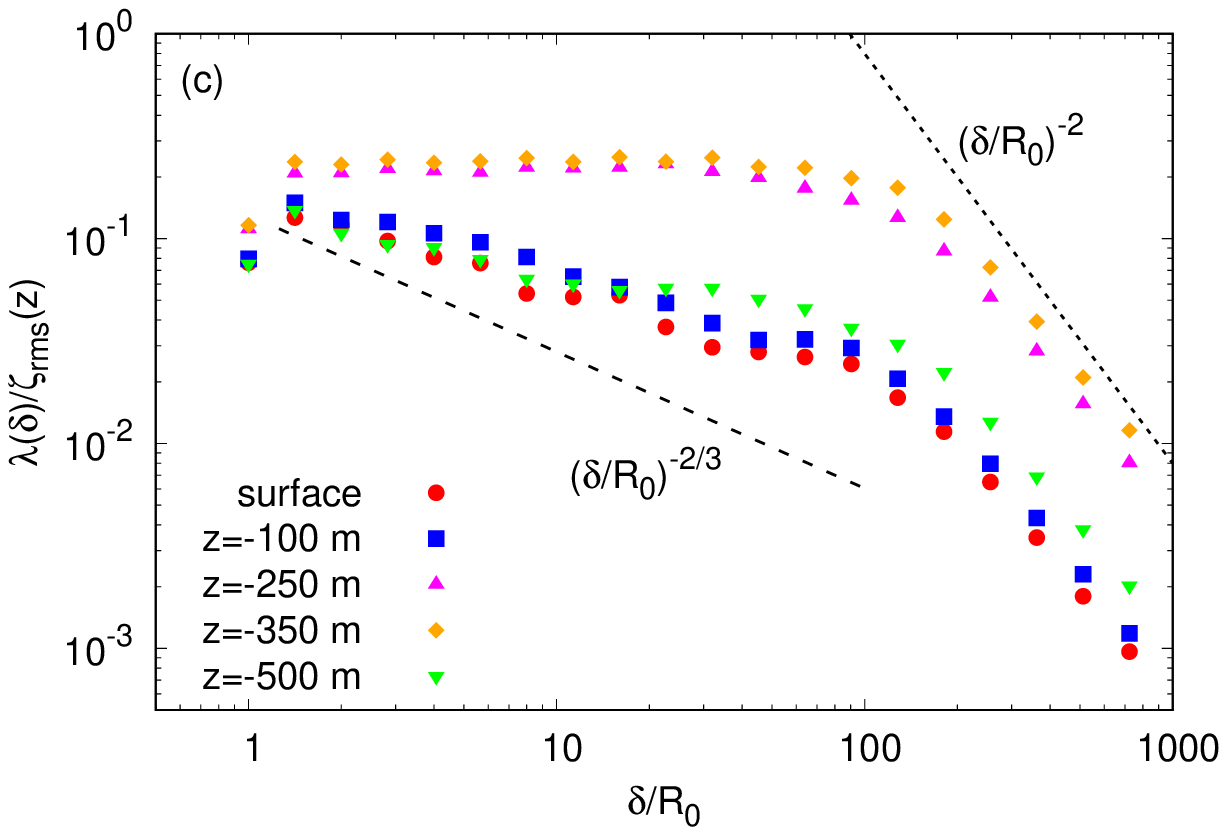}
  \caption{\label{fig:f8} FSLE-I, normalized by the rms vorticity $\zeta_\mathrm{rms}$
  as a function of the separation distance normalized by its initial value, $\delta/R_0$, 
  at the reference depths for the TC (a), ML (b) and F (c) cases.} 
\end{figure*}
%%%%%%%%%%%%%%%%%%%%%%%%%%%%

Let us now consider dispersion indicators at fixed length scale, which are less affected by the superposition 
of different regimes associated with particle pairs having different separation distances at the same time. 
We first consider relative diffusivity, defined as
\begin{equation}
\kappa = \frac{1}{2} \frac{\mathrm{d} \langle R^2(t) \rangle}{\mathrm{d}t}.
\label{eq:reldiff}
\end{equation}
This is presented in Fig.~\ref{fig:f7}, as a function of the separation distance $\delta=\langle R^2(t) \rangle^{1/2}$, 
after rescaling time by $1/\zeta_\mathrm{rms}$ and distance by $R_0$. This diagnostic returns a picture more in adequacy with the theoretical 
expectations based on the shape of the kinetic energy spectrum. 
Even if in some cases the curves show some wiggles, on average we find that the scale-by-scale relative diffusivity 
reasonably scales according to the dimensionally expected behaviors, $\delta^2$ (corresponding to a spectral exponent 
$\beta>3$ and nonlocal dispersion) and $\delta^{4/3}$ (corresponding to $\beta=5/3$ and local dispersion).
The first scaling behavior is observed in the interior for the TC case, 
below the mixed layer in the ML case, and in the upper thermocline for the F case. 
At the surface (for all cases), at the base of the mixed layer
(for the ML and F cases) and at the bottom (for the TC and F cases), instead, the results support 
the second scaling behavior.
The quite good collapse of data from different models and at different depths (in the same scale ranges as those from
relative dispersion) onto general behaviors 
determined by the kinetic energy spectrum, provides 
a first clear evidence of universal dispersion regimes controlled by the dynamical properties of the 
turbulent flows. 

The dispersion rate at fixed length scale is quantified by the FSLE. Since here we consider 
the separation process of two particles advected by the same flow starting from different positions, 
we refer to the FSLE-I, which is computed as
\begin{equation}
\lambda(\delta) = \frac{\log{r}}{\langle \tau(\delta) \rangle},
\label{eq:fsle_r}
\end{equation}
where the average is over all pairs and $\tau(\delta)$ is the time needed to observe the growth of separation from a
scale $\delta$ to a scale $r\delta$ (with $r > 1$).
We verified that the results do not appreciably change when using 
a generalization of the previous definition to discrete time~\cite{ABCCV1997,CV2013}. 
The amplification factor was set to $r=\sqrt{2}$, but we checked 
the robustness of the results with respect to this choice. 

Let us recall that, dimensionally, one expects the FSLE to scale as
$\lambda(\delta) \sim \delta^{(\beta-3)/2}$ for a kinetic energy spectrum 
$E(k) \sim k^{-\beta}$ (see Refs.~\onlinecite{LaCasce2008,FBPL2017}). 
If $\beta<3$ one then has a power-law behavior of the FSLE, corresponding to a local dispersion process.
In the case of Kolmogorov scaling, $\beta=5/3$ and hence $\lambda(\delta) \sim \delta^{-2/3}$, a behavior that is directly 
related to Richardson superdiffusive regime, $\langle R^2(t) \rangle \sim t^3$. 
When $\beta>3$, instead, \ie~when the advecting flow is smooth, the FSLE is expected to be constant, which indicates
a nonlocal dispersion regime. 
Finally, at scales much larger than the largest eddies, the FSLE has a diffusive scaling $\lambda(\delta) \sim \delta^{-2}$.

Figure~\ref{fig:f8} reports the FSLE-I, rescaled by $\zeta_\mathrm{rms}$, as a function of the separation distance 
rescaled by its initial value, $\delta/R_0$. The results quite clearly indicate that dispersion is nonlocal 
(constant FSLE-I, with $\lambda(\delta) \approx 0.2 \zeta_\mathrm{rms}$) 
over a broad range of scales up to $O(100) R_0$, in the interior of the TC system,
as well as below the mixed layer in the ML case. It is also the case for the F model, provided $|z|$ is not too large.
Close to the vertical boundaries and in the mixed layer, when present, dispersion is instead always local.
The value of $\lambda(\delta)$ at the smallest separations, which should provide an estimate of $\lambda_L$, is found to be
quite close to $0.1 \zeta_\mathrm{rms}$.
In the range $\delta<O(100) R_0$, extending up to the largest active flow scales, 
the FSLE-I displays a power-law dependence on the separation distance compatible with $\delta^{-2/3}$ 
(the theoretical expectation for Richardson superdiffusion), at least on average.  
Finally at scales larger than $O(100) R_0$, $\lambda(\delta) \sim \delta^{-2}$ in all cases, indicating a diffusive 
behavior in this range.

Summarizing, once properly rescaled with the rms vorticity, relative dispersion, relative diffusivity and the FSLE-I 
return a coherent picture that allows to identify different dispersion regimes 
and to relate them with the statistical 
features of the turbulent flows. 
In particular, the analysis reveals a transition of behavior with depth. The dispersion process is found to be local 
at the surface, while it becomes nonlocal at depth, due to the decay of small eddies.
However, while this occurs rapidly with increasing depth in the absence of mixed layer instabilities, when the latter are present
the transition is moved to larger depth, below the mixed layer, due to energetic submesoscale dynamics in the whole mixed layer.

%%%%%%%%%%%%%%%%%%%%%%%%%%%%%%%%%%%%%%%%%%%%
\subsection{\label{sec:fsle-ii} Vertical correlation of horizontal dispersion properties}
\begin{figure}
%  \centering
  \includegraphics[width=0.9\columnwidth]{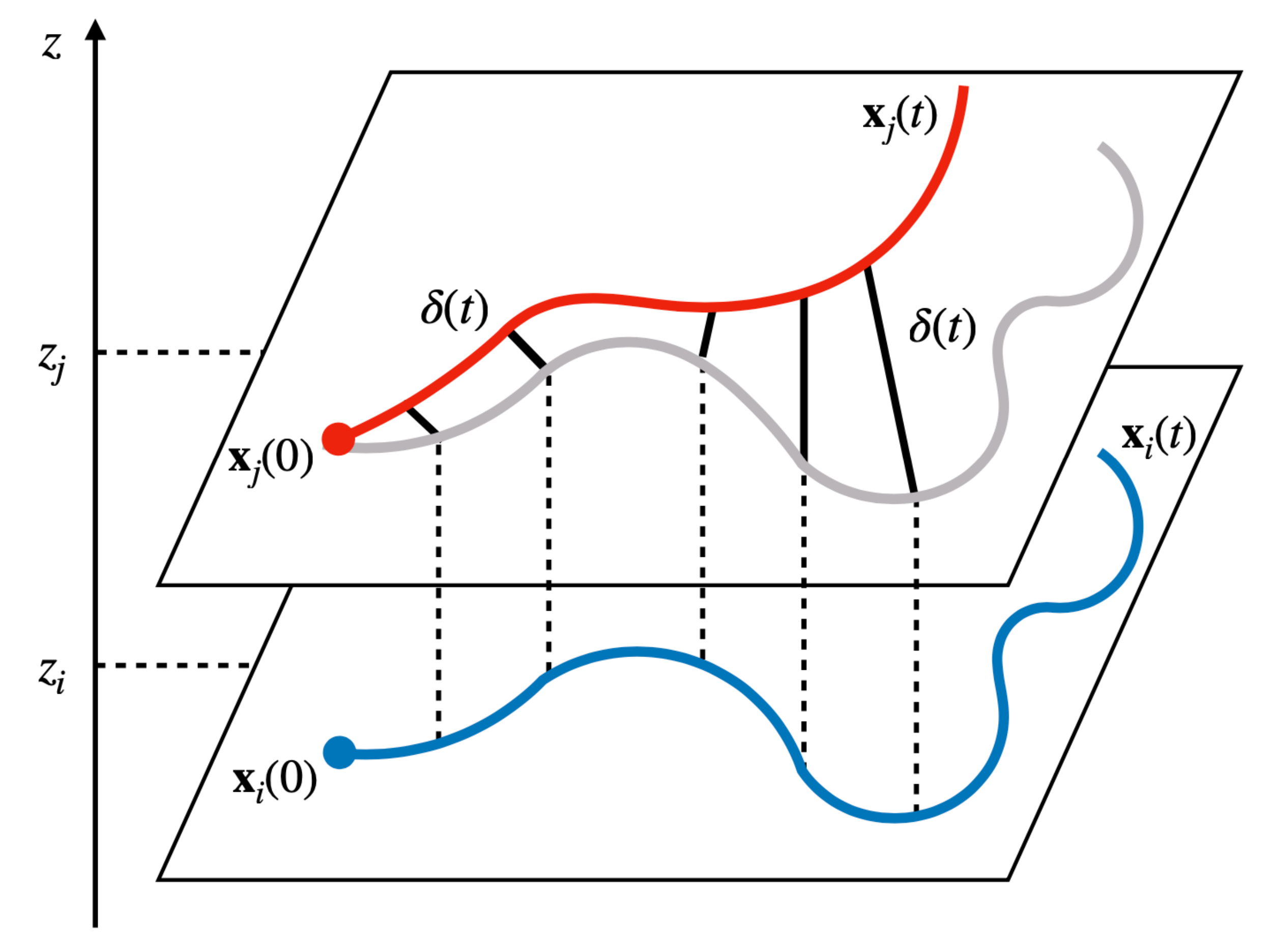}
  \caption{\label{fig:f9} Schematic illustration of typical trajectory pairs used for the computation of the FSLE-II. 
  Here $\delta(t)$ is the horizontal separation distance between trajectories $\bm{x_i}(t)$ and $\bm{x}_j(t)$, evolving on different 
  depth levels ($z_i$ and $z_j$, respectively). The gray curve is the projection of the trajectory at depth on the upper level. 
  The initial position of particles $i$ and $j$ is the same on the horizontal and is here indicated by a dot.}
\end{figure}

In the previous section, horizontal dispersion properties were discussed, depth by depth, in particular using the 
$1^\mathrm{st}$-kind FSLE. We are now interested in the relative motion between particles seeded at different depths. 
This amounts to considering the evolution of pairs of trajectories starting from the same initial position, 
on the horizontal, but with dynamics governed by different flows (see Fig.~\ref{fig:f9}).  
Their spreading process can be examined using a modified type of FSLE, as proposed in the context of predictability 
studies~\cite{ILRSV2002,LCFS2019}.

To take this approach, we consider the positions at time $t$, $\bm{x}_i(t)$ and $\bm{x}_j(t)$, of particles 
initialized on different levels ($z_i$ and $z_j$) and advected by the horizontal flow at their depth: 
\begin{eqnarray}
\frac{d\bm{x}_i}{dt} & = & \bm{v}(\bm{x}_i(t),z_i,t),\nonumber \\ 
\frac{d\bm{x}_j}{dt} & = & \bm{v}(\bm{x}_j(t),z_j,t), \nonumber
\end{eqnarray}
with 
$\bm{x}_i(0)=\bm{x}_j(0)$. 
We can still define  
as $\delta=|\bm{x}_i(t)-\bm{x}_j(t)|$ 
the horizontal separation distance for pair $(i,j)$ , 
\ie~as if the two particles were at the same level
(or, in other terms, by projecting $\bm{x}_i(t)$ on the plane $z=z_j$, as in Fig.~\ref{fig:f9}). 
We then introduce the FSLE-II with a definition analogous to that of the FSLE-I, Eq.~(\ref{eq:fsle_r}), 
and denote it $\lambda_v(\delta)$. In the following, we will always consider that one particle is at the 
surface (hence, \eg, $z_j=0$).
%%%%%%%%%%%%%%%%%%%%%%%%%%%%
% Fig. FLSE-II 
% scheme
%%%%%%%%%%%%%%%%%%%%%%%%%%%%

Before looking at the results, it is useful to discuss the behaviors expected for this indicator.
We first remark that in our case, below the surface, the turbulent intensity can be considerably lower, depending on the depth 
and the model dynamics (see the behavior of $u_\mathrm{rms}$ with $|z|$ in Fig.~\ref{fig:f4}), and vertical shear can play a relevant role. 
It is not difficult to obtain from dimensional arguments that, due to the mean shear,  
$\lambda_v(\delta) \simeq \log{(r)} \Lambda |z| \delta^{-1}$ (recall that the mean shear is $\Lambda=\Lambda_m=\Lambda_t$ 
in the present simulations, see table~\ref{tab:t1}).
This type of contribution to the FSLE-II can be expected to be large where $U(z)/u_\mathrm{rms}$ is large (see inset in Fig.~\ref{fig:f4}). 
A similar scaling of $\lambda_v(\delta)$ can also arise, more generally, from  
the shear due to the typical difference (versus depth) of the total velocity 
$\langle |\bm{v}(z)-\bm{v}(0)|^2 \rangle^{1/2}$, where $\bm{v}(z)=\bm{U}(z)+\bm{u}(x,y,z,t)$ 
(see also Eq.~(\ref{eq:lagr_motion})), from which one would expect 
$\lambda_v(\delta) \simeq \log{(r)} \langle |\bm{v}(z)-\bm{v}(0)|^2 \rangle^{1/2} \delta^{-1}$. 

Another point to bear in mind is that, in all (TC, ML, F) cases, also the small-scale energetic content of our flows is reduced 
in the upper thermocline, or below the mixed layer. 
This situation is close to the one discussed in Ref.~\onlinecite{LCFS2019}, which considers the separation of two particles 
advected by two flow fields $\bm{v}$ and $\bm{v}'$ that have identical energy spectra at large scales but one of which has no scales 
smaller than a cut-off length $\ell^*$. In such a case, the distance between the two corresponding trajectories should be 
$|\bm{x}(t)-\bm{x}'(t)| \sim |\bm{v}-\bm{v}'| t \sim v^* t$, where 
$v^*$ is the typical velocity difference, as long as $|\bm{x}(t)-\bm{x}'(t)|<\ell^*$. Thus, dimensionally, one has that 
the FSLE-II should scale as $\lambda_v(\delta) \sim \delta^{-1}$ for $\delta$ small enough. At larger scales, 
the difference between the two flows has no more influence and the FSLE-II typically recovers the behavior of the FSLE-I $\lambda(\delta)$. 
In our case, an analogous reasoning (with $\bm{v}$ and $\bm{v}'$ the velocities at the surface and at depth, respectively) would imply that 
a critical length scale $\ell^*$ should mark the transition between the behaviors $\lambda_v(\delta) \sim \delta^{-1}$ 
and $\lambda_v(\delta) \sim \lambda(\delta)$, provided the previously discussed shear contribution is weak enough. 

%%%%%%%%%%%%%%%%%%%%%%%%%%%%
% Fig. FLSE-II 
% versus depth: 
% TC, ML, F 
\begin{figure*}
%  \centering
  \includegraphics[width=0.33\textwidth]{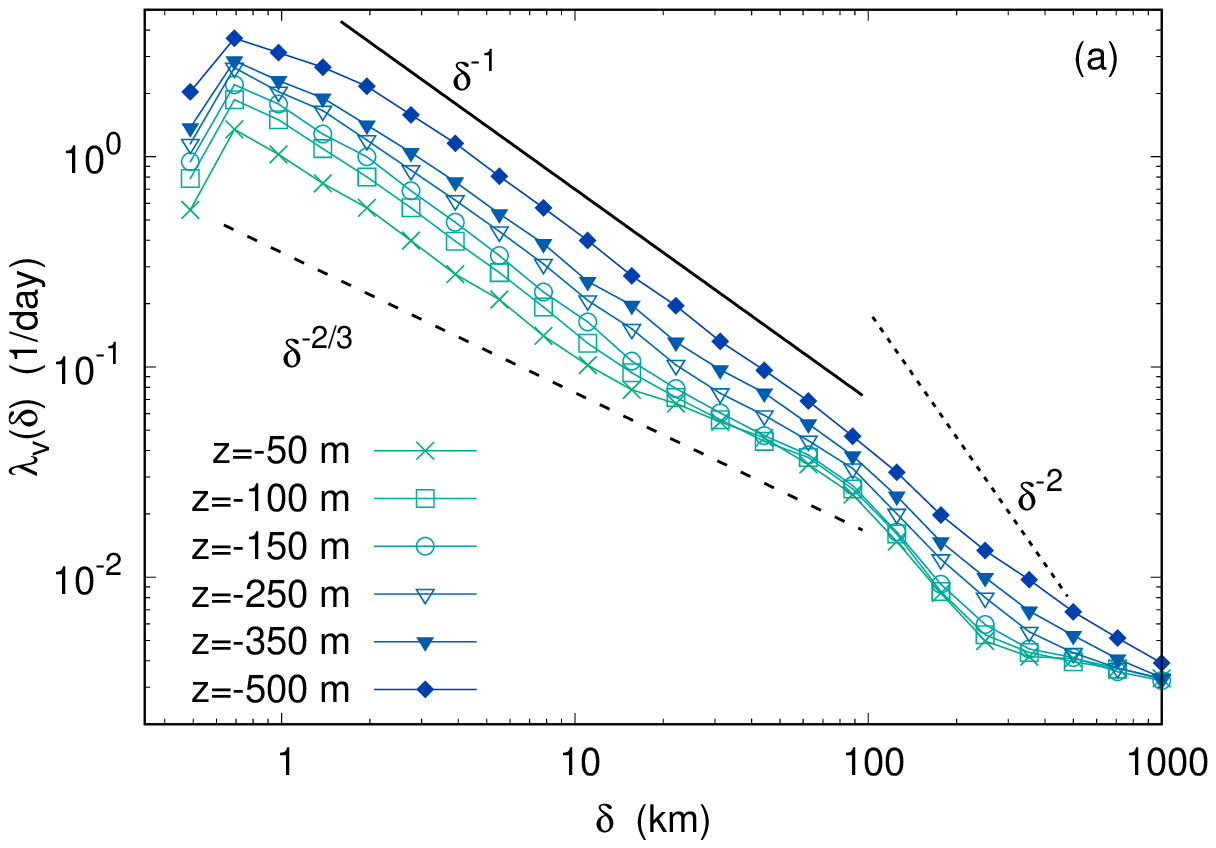}
  \includegraphics[width=0.33\textwidth]{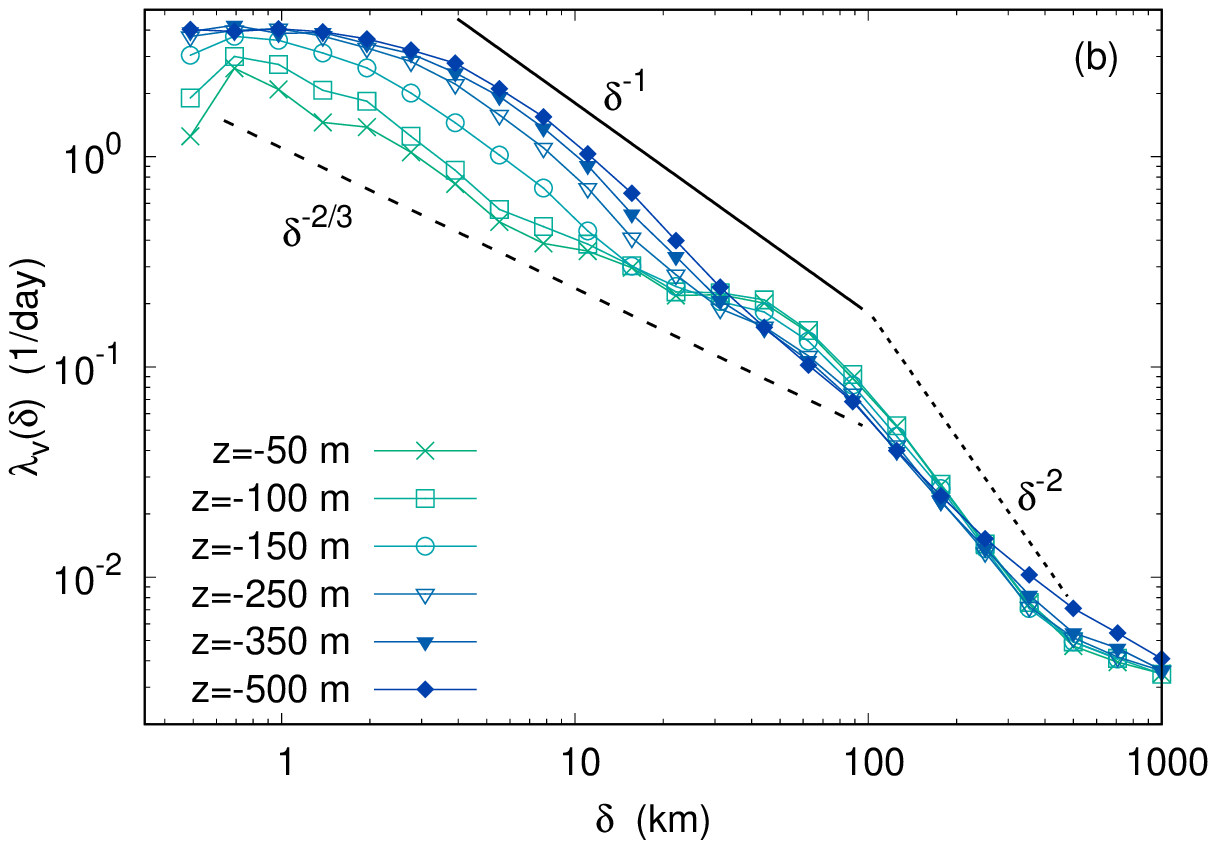}
  \includegraphics[width=0.33\textwidth]{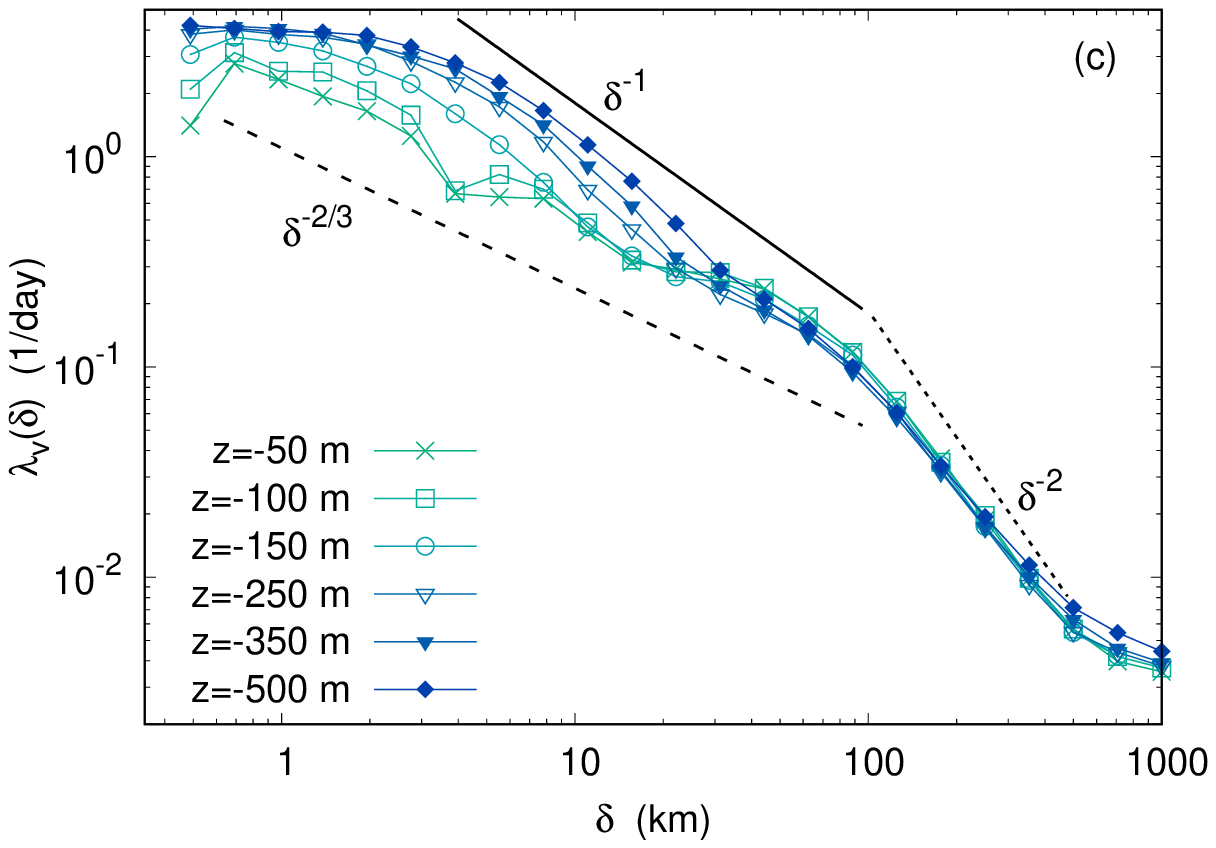}
  \caption{\label{fig:f10} FSLE-II for the TC (a), ML (b) and F (c) cases.
  The settings are as in Fig.~\ref{fig:f8} but particles in a pair are now selected such that, initially,
  the first one is at the surface and the second one is at a depth $z$, with no horizontal separation between them.
  Two additional depths, with respect to the computation of the FSLE-I, are here shown: $z=-50$~m and $z=-150$~m. 
  Darker symbols correspond to larger depths.}
\end{figure*}
%%%%%%%%%%%%%%%%%%%%%%%%%%%%
%%%%%%%%%%%%%%%%%%%%%%%%%%%%
% Fig. FLSE-II - compensated 
% versus depth: 
% TC, ML, F 
\begin{figure*}
%  \centering
  \includegraphics[width=0.33\textwidth]{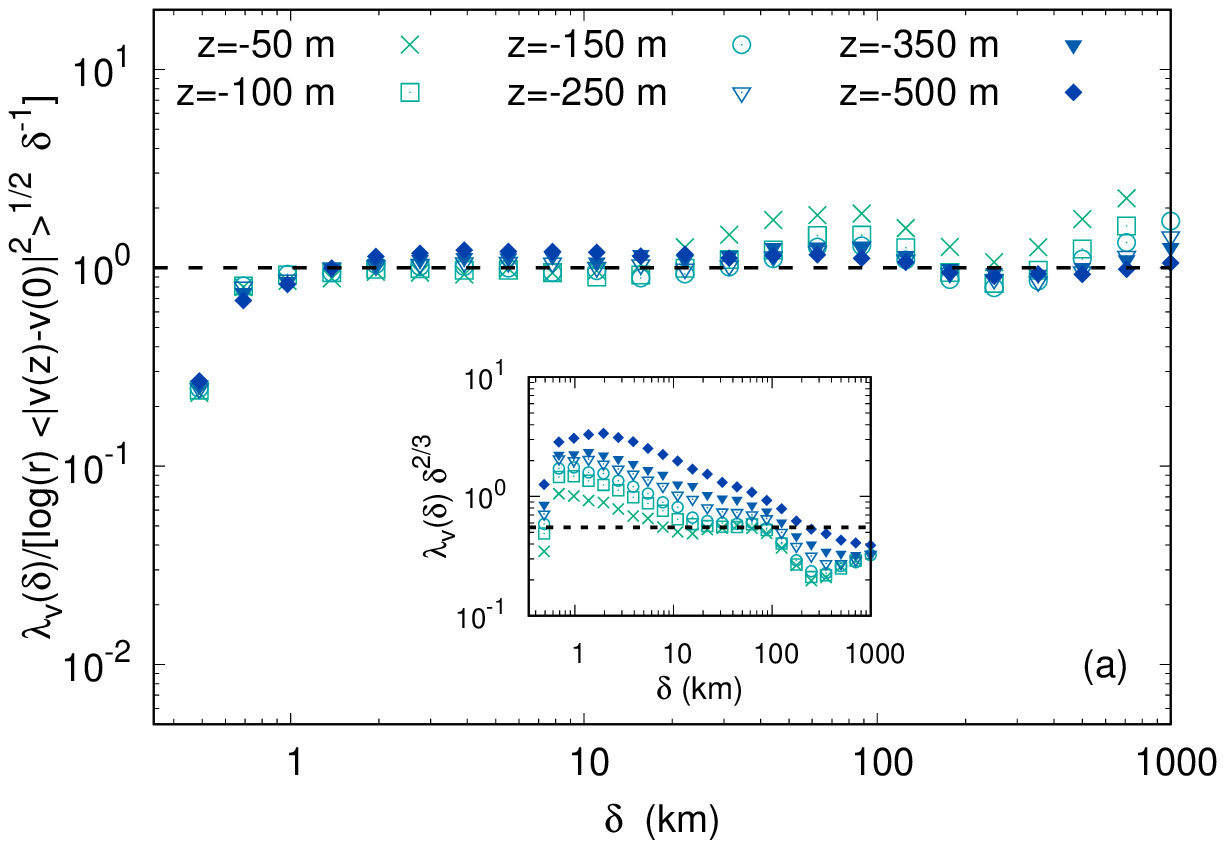}
  \includegraphics[width=0.33\textwidth]{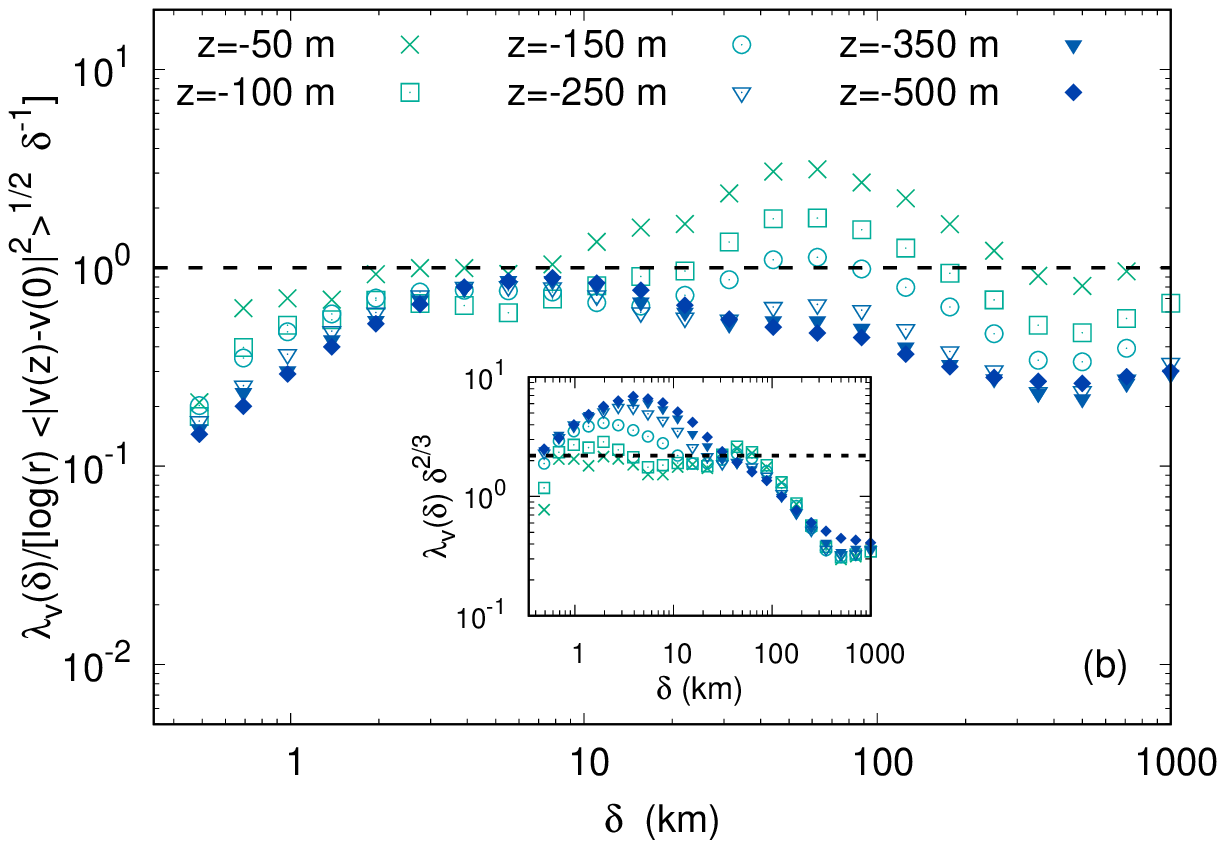}
  \includegraphics[width=0.33\textwidth]{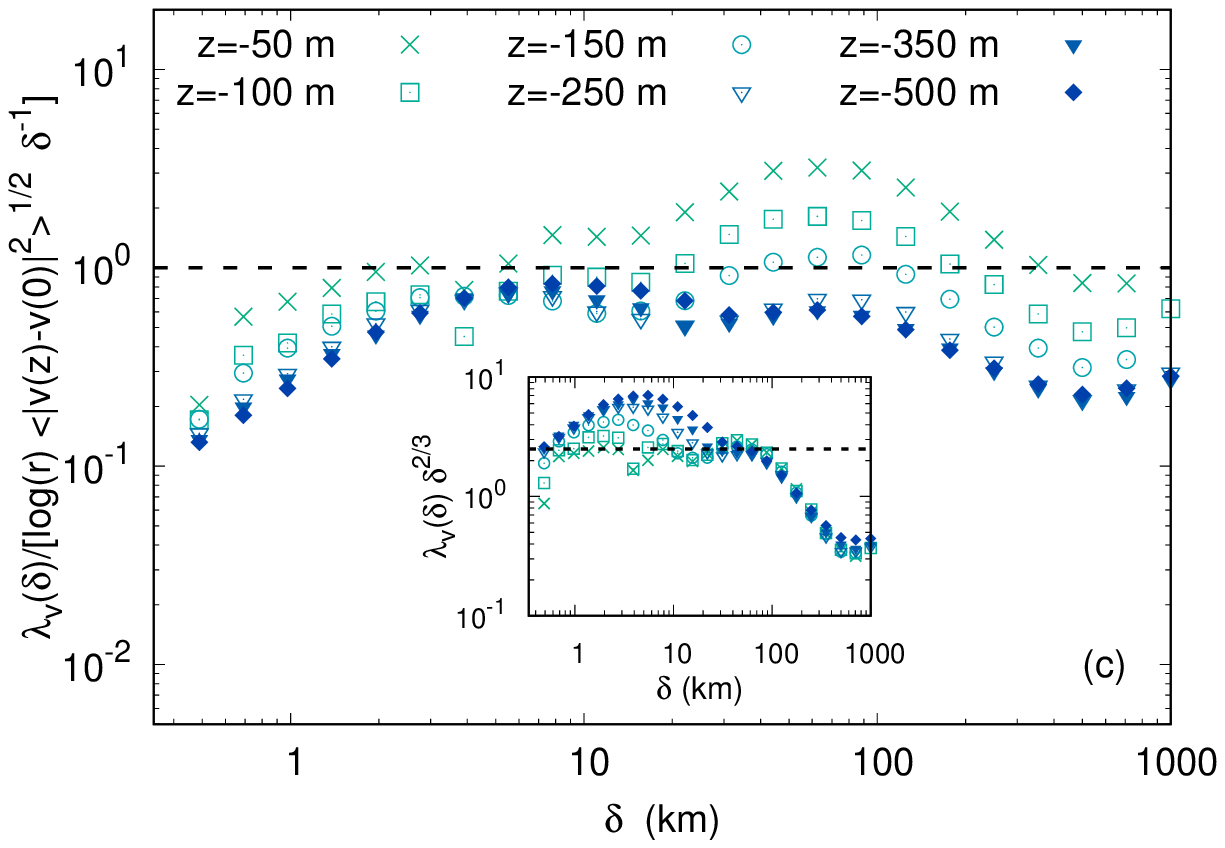}
  \caption{\label{fig:f11} FSLE-II for the TC (a), ML (b) and F (c) cases; darker symbols correspond to larger depths. 
  In the main panels $\lambda_v(\delta)$ is compensated by the expectation in the presence of the typical vertical shear due to both 
  the mean flow and turbulent fluctuations, $\log{(r)} \langle |\bm{v}(z)-\bm{v}(0)|^2 \rangle^{1/2} \delta^{-1}$ 
  (where $\bm{v}(x,y,z,t)$ is the total velocity at depth $z$). The insets show $\lambda_v(\delta)$ compensated by 
  $\delta^{-2/3}$; the dashed lines are guides to the eye for constant values.} 
\end{figure*}
%%%%%%%%%%%%%%%%%%%%%%%%%%%%
Let us now illustrate the results, which are obtained using $4096$ original pairs, as for the computation of the FSLE-I 
(but now selecting one particle at the surface and another one at depth). 
Our interest is mainly focused on the scale range between $O(1)$~km and $O(100)$~km.  
The FSLE-II is shown in Fig.~\ref{fig:f10} for the reference depths and two additional ones, $z=-50$~m and $z=-150$~m, 
respectively above and below the mixed layer, when present. 
At small depths the slope of the FSLE-II 
is close to that of the FSLE-I (recall that $\lambda(\delta) \sim \delta^{-2/3}$ at the surface, before the onset 
of the diffusive regime $\delta^{-2}$ at the largest values of $\delta$). 
This is particularly evident for the ML and F cases, where the behavior $\sim \delta^{-2/3}$ is observed over a broader 
range of separations, namely from few to slightly less than $100$~km. 
For the TC model, already quite close to the surface ($z=-50$~m in Fig.~\ref{fig:f10}a), at separations $\delta < O(10)$~km, 
however, the slope of the FSLE-II gets definitely larger in absolute value ($\lambda_v(\delta) \sim \delta^{-1}$). As the vertical shear is still weak 
at such depths, it is possible to associate this change of scaling with the missing small scales in the deeper flow. 
It also appears reasonable, here, that the crossover scale is $\ell^* \approx 10$~km, as this value is also close to the length scale of  
the mixed-layer instability (capable of energizing the submesoscale in the first $100$~m below the surface), 
which is absent in this model. A similar transition to $\lambda_v(\delta) \sim \delta^{-1}$ becomes evident only below the mixed layer 
($z=-150$~m in Figs.~\ref{fig:f10}b,c) in the ML and F cases, for which small scales are energetic down to $z=-100$~m.  
Further below the surface, the FSLE-II is in all cases close to $\delta^{-1}$ (though slightly steeper for the ML and F models, where 
$\lambda_v(\delta) \sim \delta^{-\alpha}$, with $\alpha \simeq 1.25$ and $1.15$, respectively) over a more extended range 
of separations, due to the missing small scales, but now also due to the vertical shear becoming more important with depth and 
eventually dominating. 
Interestingly, some indications about the relevant role of vertical shear were recently 
documented also in a more realistic, albeit more specific, numerical study addressing pair dispersion 
at submesoscales in the Bay of Bengal~\cite{Essink2019}. 
At the largest depths, a flattening of the FSLE-II at the smallest separations is seen, particularly for the ML and F cases.
It should be noted, however, that in this range of scales and depths, due to the large velocity differences involved, 
the results may be affected by the finite temporal resolution of the data.  

To better appreciate the contribution from the vertical shear, in Fig.~\ref{fig:f11} we report $\lambda_v(\delta)$
compensated by the expectation in the presence of
the vertical shear arising from both the mean flow and the turbulent velocity,
$\log{(r)} \langle |\bm{v}(z) - \bm{v}(0)|^2 \rangle^{1/2} \delta^{-1}$ (main panels).
The compensation by the contribution from the mean vertical shear only, \ie~$\log{(r)} \Lambda|z|\delta^{-1}$,
was not found to be sufficient to account for the behavior of the FSLE-II (not shown).
On the contrary, the FSLE-II compensated by the total-shear prediction approaches the constant value $1$
(Fig.~\ref{fig:f11}), particularly at larger depths.
Note that the collapse of the different curves corresponding to different depths is much better for the TC case
than for the ML and F cases.
For the TC case, the deviations at the smallest depths (e.g. $z=-50$~m) for separations $\delta>10$~km are related to the
scaling $\lambda_v(\delta) \sim \delta^{-2/3}$, as shown by the inset of Fig.~\ref{fig:f11}a.
Indeed, due to still energetic eddies in both flows at these scales, and weak vertical shear at these depths, as already
observed from Fig.~\ref{fig:f10}a, in this range the FSLE-II is close to the FSLE-I.
In the ML and F cases, when depths larger than the mixed-layer depth ($h=100$~m) are considered,
the compensated (by $\delta^{-1}$) FSLE-II is fairly close to $1$ in a broad range of horizontal separations
($2$~km~$\leq \delta \leq 100$~km).
More important deviations are observed at smaller depths (inside the mixed layer) and can be attributed
to the scaling $\lambda_v(\delta) \approx \delta^{-2/3}$
(insets of Fig.\ref{fig:f11}b and c), related to the similar small-scale energetic content of the flows at the surface
and below it, in this range of depths.

%%%%%%%%%%%%%%%%%%%%%%%%%%%%%%%%%%%%%%%%%%%%
\section{\label{sec:concl} Conclusions}
We explored Lagrangian pair dispersion in stratified upper-ocean turbulence. 
We focused on the identification of different dispersion regimes and on the 
possibility to relate the characteristics of the spreading process at the surface and at depth. 
The latter question is particularly relevant to assess the possibility of inferring the 
dynamical features of deeper flows from the experimentally more accessible (\eg~by satellite altimetry) surface ones. 
In this sense, Lagrangian dispersion statistics can provide useful information to understand the coupling between 
the surface and interior dynamics. 
Some perspectives on the use of further particle-based approaches to this subject, 
which is key to understand how submesoscale flows participate in biogeochemical and heat budgets, 
are discussed in Ref.~\onlinecite{Mahadevan_etal_2020}.

Tracer particles were advected by turbulent flows characterized by energetic submesoscales close to the surface, 
both in the presence (ML and F cases) and in the absence (TC case) of mixed-layer instabilities. 
The numerical simulations of the model dynamics~\cite{CFFF2016} were carried out using realistic parameter values 
for the midlatitude ocean. 
Even if the presence of a mixed layer has a signature at the surface in terms of a less filamentary flow field, 
its main effect is to energize the full upper part of the water column and, hence, to strongly impact the vertical 
variation of the statistical features of turbulence. Kinetic energy spectra close to $k^{-5/3}$ are found at the 
surface in all models. They are instead steeper than $k^{-3}$ at depth, due to the decay of small eddies. 
However, while this change of behavior occurs already close to the surface in the TC case, in the ML and F cases, 
it only manifests below the depth of the mixed layer ($z=-100$~m). 

The different statistical indicators examined, once properly rescaled to take into account the typical intensity 
of velocity gradients, allowed to group the data at different depths and from different models into only two 
universal behaviors, corresponding to nonlocal and local dispersion, which are in agreement with the 
dimensional expectations based on kinetic energy spectra.  
Therefore, our results indicate a clear transition of dispersion regime with depth, which is quite generic. 
The spreading process is local at the surface. In the absence of a mixed layer it very soon changes to nonlocal 
at small depths, while in the opposite case this only occurs at larger depths, below the mixed layer.

It is here worth commenting on our results from a dimensional point of view. Horizontal dispersion is always found to be 
diffusive-like at spatial scales larger than $O(100)$~km and at times $t>30$~day from the release. The intensity of the 
dispersion process, as quantified, \eg, by the FSLE-I, decreases with depth in all models, except close to the bottom 
boundary in the TC and F cases (due the small-scale flows gaining energy again there). 
In the nonlocal-dispersion cases, the flat behavior of the FSLE-I in a broad range of scales 
($O(1)$~km$<\delta<O(100)$~km) allows to estimate the Lyapunov exponent $\lambda_L$. The latter is found to be 
of order $0.01$~day$^{-1}$ in the TC case and $0.1$~day$^{-1}$ in the ML and F ones. 
Under local-dispersion, the behaviors found are compatible with Richardson superdiffusion from few km to 
about $100$~km. The scale-by-scale dispersion rate, $\lambda(\delta)$, is considerably enhanced at submesoscales, 
reaching values $\approx (0.1-0.2)$~day$^{-1}$ in the TC case and $\approx 1$~day$^{-1}$, 
compatible with surface-drifter observations in different regions~\cite{CLPSZ2017}, in the ML, F cases. 

We further investigated the transition from local to nonlocal dispersion, with increasing depth, by means of the FSLE-II. 
Our results indicate that, in the absence of a mixed layer, dispersion properties rapidly decorrelate from those at the surface.
In the ML and F cases, instead, a similar phenomenon occurs only below the mixed layer.
The transition is sharper in the TC model. However, the relation between dispersion at depth and at the surface 
appears in this case to be largely controlled by the vertical shear (due to the total velocity), as revealed by the
very good collapse of the data rescaled by the prediction based on it. This suggests that, even in this case, 
it should be in principle possible to infer how a tracer at depth separates from one at the surface, if the shear is known, or 
to parameterize it using information at the surface only.
In the presence of mixed-layer instabilities, the results indicate  
that the statistical properties of the spreading process at the surface can be considered as a good proxy 
of those in the whole mixed layer. 
The vertical-shear prediction for the FSLE-II still appears reasonable, 
particularly below the mixed layer and for separations ranging from few km to $\approx 100$~km,
but now the agreement is essentially limited to the order of magnitude, 
which makes it more difficult to establish a link between 
the interior and surface dispersion. 

Finally, based on the above considerations, in our opinion, this study provides evidence of the interest 
for future satellite altimetry, as the SWOT mission~\cite{Morrow_etal_2019}, that should provide surface velocity fields 
at unprecedented high resolution, also in the light of understanding subsurface ocean dynamics.  

%%%%%%%%%%%%%%%%%%%%%%%%%%%%%%%%%%%%%%%%%%%%
\vspace{-0.2cm}

\begin{acknowledgments}
This work is a contribution to the joint CNES-NASA SWOT projects ``New dynamical tools'' and DIEGO
and is supported by the French CNES TOSCA program.
\end{acknowledgments}

%%%%%%%%%%%%%%%%%%%%%%%%%%%%%%%%%%%%%%%%%%%%
\appendix*
%%%%%%%%%%%%%%%%%%%%%%%%%%%%%%%%%%%%%%%%%%%%
\section{\label{sec:appA} Streamfunction at arbitrary depth}
The streamfunction $\psi$ at a generic depth $z$ can be expressed, in Fourier space (with $\bm{k}$ the horizontal 
wavenumber) in terms of $\psi_1$, $\psi_2$, $\psi_3$ as
\begin{eqnarray}
\hat{\psi}(\bm{k},z) = \frac{1}{\sinh{\mu_m}} \left[ \hat{\psi}_1(\bm{k}) \sinh \left( \mu_m \frac{z+h}{h} \right)  - \right.  & \nonumber\\
- \left. \hat{\psi}_2(\bm{k}) \sinh \left( \mu_m \frac{z}{h} \right) \right] & \quad 
\label{eq:psi_z1}
\end{eqnarray}
in layer $1$ ($-h < z \leq 0$) and as
\begin{eqnarray}
\hat{\psi}(\bm{k},z) = \frac{1}{\sinh{\mu_t}} \left[ \hat{\psi}_2(\bm{k}) \sinh \left( \frac{N_t}{N_m} \mu_m \frac{z+H}{h} \right) 
- \right.  & \nonumber \\
- \left. \hat{\psi}_3(\bm{k}) \sinh \left( \frac{N_t}{N_m} \mu_m \frac{z+h}{h} \right) \right]  & \quad
\label{eq:psi_z2}
\end{eqnarray}
in layer $2$ ($-H \leq z \leq -h$).

%%%%%%%%%%%%%%%%%%%%%%%%%%%%%%%%%%%%%%%%%%%%
% references
\section*{References}
\bibliography{references}

\providecommand{\noopsort}[1]{}\providecommand{\singleletter}[1]{#1}%
\begin{thebibliography}{57}%
\makeatletter
\providecommand \@ifxundefined [1]{%
 \@ifx{#1\undefined}
}%
\providecommand \@ifnum [1]{%
 \ifnum #1\expandafter \@firstoftwo
 \else \expandafter \@secondoftwo
 \fi
}%
\providecommand \@ifx [1]{%
 \ifx #1\expandafter \@firstoftwo
 \else \expandafter \@secondoftwo
 \fi
}%
\providecommand \natexlab [1]{#1}%
\providecommand \enquote  [1]{``#1''}%
\providecommand \bibnamefont  [1]{#1}%
\providecommand \bibfnamefont [1]{#1}%
\providecommand \citenamefont [1]{#1}%
\providecommand \href@noop [0]{\@secondoftwo}%
\providecommand \href [0]{\begingroup \@sanitize@url \@href}%
\providecommand \@href[1]{\@@startlink{#1}\@@href}%
\providecommand \@@href[1]{\endgroup#1\@@endlink}%
\providecommand \@sanitize@url [0]{\catcode `\\12\catcode `\$12\catcode
  `\&12\catcode `\#12\catcode `\^12\catcode `\_12\catcode `\%12\relax}%
\providecommand \@@startlink[1]{}%
\providecommand \@@endlink[0]{}%
\providecommand \url  [0]{\begingroup\@sanitize@url \@url }%
\providecommand \@url [1]{\endgroup\@href {#1}{\urlprefix }}%
\providecommand \urlprefix  [0]{URL }%
\providecommand \Eprint [0]{\href }%
\providecommand \doibase [0]{http://dx.doi.org/}%
\providecommand \selectlanguage [0]{\@gobble}%
\providecommand \bibinfo  [0]{\@secondoftwo}%
\providecommand \bibfield  [0]{\@secondoftwo}%
\providecommand \translation [1]{[#1]}%
\providecommand \BibitemOpen [0]{}%
\providecommand \bibitemStop [0]{}%
\providecommand \bibitemNoStop [0]{.\EOS\space}%
\providecommand \EOS [0]{\spacefactor3000\relax}%
\providecommand \BibitemShut  [1]{\csname bibitem#1\endcsname}%
\let\auto@bib@innerbib\@empty
%</preamble>
\bibitem [{\citenamefont {McWilliams}(2016)}]{McWilliams2016}%
  \BibitemOpen
  \bibfield  {author} {\bibinfo {author} {\bibfnamefont {J.~C.}\ \bibnamefont
  {McWilliams}},\ }\bibfield  {title} {\enquote {\bibinfo {title} {Submesoscale
  currents in the ocean},}\ }\href@noop {} {\bibfield  {journal} {\bibinfo
  {journal} {Proc. R. Soc. A}\ }\textbf {\bibinfo {volume} {472}},\ \bibinfo
  {pages} {20160117} (\bibinfo {year} {2016})}\BibitemShut {NoStop}%
\bibitem [{\citenamefont {L\'evy}(2008)}]{Levy2008}%
  \BibitemOpen
  \bibfield  {author} {\bibinfo {author} {\bibfnamefont {M.}~\bibnamefont
  {L\'evy}},\ }\bibfield  {title} {\enquote {\bibinfo {title} {The modulation
  of biological production by oceanic mesoscale turbulence},}\ }\href@noop {}
  {\bibfield  {journal} {\bibinfo  {journal} {Lect. Notes Phys.}\ }\textbf
  {\bibinfo {volume} {744}},\ \bibinfo {pages} {219--–261} (\bibinfo {year}
  {2008})}\BibitemShut {NoStop}%
\bibitem [{\citenamefont {Ferrari}(2011)}]{Ferrari2011}%
  \BibitemOpen
  \bibfield  {author} {\bibinfo {author} {\bibfnamefont {R.}~\bibnamefont
  {Ferrari}},\ }\bibfield  {title} {\enquote {\bibinfo {title} {A frontal
  challenge for climate models},}\ }\href@noop {} {\bibfield  {journal}
  {\bibinfo  {journal} {Science}\ }\textbf {\bibinfo {volume} {332}},\ \bibinfo
  {pages} {316–317} (\bibinfo {year} {2011})}\BibitemShut {NoStop}%
\bibitem [{\citenamefont {Su}\ \emph {et~al.}(2018)\citenamefont {Su},
  \citenamefont {Wang}, \citenamefont {Klein}, \citenamefont {Thompson},\ and\
  \citenamefont {Menemenlis}}]{Su_etal_2018}%
  \BibitemOpen
  \bibfield  {author} {\bibinfo {author} {\bibfnamefont {Z.}~\bibnamefont
  {Su}}, \bibinfo {author} {\bibfnamefont {J.}~\bibnamefont {Wang}}, \bibinfo
  {author} {\bibfnamefont {P.}~\bibnamefont {Klein}}, \bibinfo {author}
  {\bibfnamefont {A.~F.}\ \bibnamefont {Thompson}}, \ and\ \bibinfo {author}
  {\bibfnamefont {D.}~\bibnamefont {Menemenlis}},\ }\bibfield  {title}
  {\enquote {\bibinfo {title} {Ocean submesoscales as a key component of the
  global heat budget},}\ }\href@noop {} {\bibfield  {journal} {\bibinfo
  {journal} {Nat. Commun.}\ }\textbf {\bibinfo {volume} {9}},\ \bibinfo {pages}
  {775} (\bibinfo {year} {2018})}\BibitemShut {NoStop}%
\bibitem [{\citenamefont {Bracco}, \citenamefont {Liu},\ and\ \citenamefont
  {Sun}(2019)}]{Bracco_etal_2019}%
  \BibitemOpen
  \bibfield  {author} {\bibinfo {author} {\bibfnamefont {A.}~\bibnamefont
  {Bracco}}, \bibinfo {author} {\bibfnamefont {G.}~\bibnamefont {Liu}}, \ and\
  \bibinfo {author} {\bibfnamefont {D.}~\bibnamefont {Sun}},\ }\bibfield
  {title} {\enquote {\bibinfo {title} {Mesoscale-submesoscale interactions in
  the {Gulf} of {Mexico}: from oil dispersion to climate},}\ }\href@noop {}
  {\bibfield  {journal} {\bibinfo  {journal} {Chaos, Solitons and Fractals}\
  }\textbf {\bibinfo {volume} {119}},\ \bibinfo {pages} {63--72} (\bibinfo
  {year} {2019})}\BibitemShut {NoStop}%
\bibitem [{\citenamefont {Siegelman}(2020)}]{Siegelman2020}%
  \BibitemOpen
  \bibfield  {author} {\bibinfo {author} {\bibfnamefont {L.}~\bibnamefont
  {Siegelman}},\ }\bibfield  {title} {\enquote {\bibinfo {title} {Energetic
  submesoscale dynamics in the ocean interior},}\ }\href@noop {} {\bibfield
  {journal} {\bibinfo  {journal} {J. Phys. Oceanogr.}\ }\textbf {\bibinfo
  {volume} {50}},\ \bibinfo {pages} {727--749} (\bibinfo {year}
  {2020})}\BibitemShut {NoStop}%
\bibitem [{\citenamefont {Lapeyre}\ and\ \citenamefont {Klein}(2006)}]{LK2006}%
  \BibitemOpen
  \bibfield  {author} {\bibinfo {author} {\bibfnamefont {G.}~\bibnamefont
  {Lapeyre}}\ and\ \bibinfo {author} {\bibfnamefont {P.}~\bibnamefont
  {Klein}},\ }\bibfield  {title} {\enquote {\bibinfo {title} {Dynamics of the
  upper oceanic layers in terms of surface quasigeostrophy theory},}\
  }\href@noop {} {\bibfield  {journal} {\bibinfo  {journal} {J. Phys.
  Oceanogr.}\ }\textbf {\bibinfo {volume} {36}},\ \bibinfo {pages} {165--176}
  (\bibinfo {year} {2006})}\BibitemShut {NoStop}%
\bibitem [{\citenamefont {Klein}\ \emph {et~al.}(2011)\citenamefont {Klein},
  \citenamefont {Lapeyre}, \citenamefont {Roullet}, \citenamefont {Gentil},\
  and\ \citenamefont {Sasaki}}]{Klein_etal_2011}%
  \BibitemOpen
  \bibfield  {author} {\bibinfo {author} {\bibfnamefont {P.}~\bibnamefont
  {Klein}}, \bibinfo {author} {\bibfnamefont {G.}~\bibnamefont {Lapeyre}},
  \bibinfo {author} {\bibfnamefont {G.}~\bibnamefont {Roullet}}, \bibinfo
  {author} {\bibfnamefont {S.~L.}\ \bibnamefont {Gentil}}, \ and\ \bibinfo
  {author} {\bibfnamefont {H.}~\bibnamefont {Sasaki}},\ }\bibfield  {title}
  {\enquote {\bibinfo {title} {Ocean turbulence at meso and submesoscales:
  connection between surface and interior dynamics},}\ }\href@noop {}
  {\bibfield  {journal} {\bibinfo  {journal} {Geophys. Astrophys. Fluid Dyn.}\
  }\textbf {\bibinfo {volume} {105}},\ \bibinfo {pages} {421--437} (\bibinfo
  {year} {2011})}\BibitemShut {NoStop}%
\bibitem [{\citenamefont {Roullet}\ \emph {et~al.}(2012)\citenamefont
  {Roullet}, \citenamefont {McWilliams}, \citenamefont {Capet},\ and\
  \citenamefont {Molemaker}}]{RMCM2012}%
  \BibitemOpen
  \bibfield  {author} {\bibinfo {author} {\bibfnamefont {G.}~\bibnamefont
  {Roullet}}, \bibinfo {author} {\bibfnamefont {J.~C.}\ \bibnamefont
  {McWilliams}}, \bibinfo {author} {\bibfnamefont {X.}~\bibnamefont {Capet}}, \
  and\ \bibinfo {author} {\bibfnamefont {M.~J.}\ \bibnamefont {Molemaker}},\
  }\bibfield  {title} {\enquote {\bibinfo {title} {Properties of steady
  geostrophic turbulence with isopycnal outcropping},}\ }\href@noop {}
  {\bibfield  {journal} {\bibinfo  {journal} {J. Phys. Oceanogr.}\ }\textbf
  {\bibinfo {volume} {42}},\ \bibinfo {pages} {18--38} (\bibinfo {year}
  {2012})}\BibitemShut {NoStop}%
\bibitem [{\citenamefont {Capet}\ \emph {et~al.}(2016)\citenamefont {Capet},
  \citenamefont {Roullet}, \citenamefont {Klein},\ and\ \citenamefont
  {Maze}}]{Capet_etal_2016}%
  \BibitemOpen
  \bibfield  {author} {\bibinfo {author} {\bibfnamefont {X.}~\bibnamefont
  {Capet}}, \bibinfo {author} {\bibfnamefont {G.}~\bibnamefont {Roullet}},
  \bibinfo {author} {\bibfnamefont {P.}~\bibnamefont {Klein}}, \ and\ \bibinfo
  {author} {\bibfnamefont {G.}~\bibnamefont {Maze}},\ }\bibfield  {title}
  {\enquote {\bibinfo {title} {Intensification of upper-ocean submesoscale
  turbulence through charney baroclinic instability},}\ }\href@noop {}
  {\bibfield  {journal} {\bibinfo  {journal} {J. Phys. Oceanogr.}\ }\textbf
  {\bibinfo {volume} {46}},\ \bibinfo {pages} {3365--3384} (\bibinfo {year}
  {2016})}\BibitemShut {NoStop}%
\bibitem [{\citenamefont {Boccaletti}, \citenamefont {Ferrari},\ and\
  \citenamefont {Fox-Kemper}(2006)}]{BFF2007}%
  \BibitemOpen
  \bibfield  {author} {\bibinfo {author} {\bibfnamefont {G.}~\bibnamefont
  {Boccaletti}}, \bibinfo {author} {\bibfnamefont {R.}~\bibnamefont {Ferrari}},
  \ and\ \bibinfo {author} {\bibfnamefont {B.}~\bibnamefont {Fox-Kemper}},\
  }\bibfield  {title} {\enquote {\bibinfo {title} {Mixed layer instabilities
  and restratification},}\ }\href@noop {} {\bibfield  {journal} {\bibinfo
  {journal} {J. Phys. Oceanogr.}\ }\textbf {\bibinfo {volume} {37}},\ \bibinfo
  {pages} {2228--2250} (\bibinfo {year} {2006})}\BibitemShut {NoStop}%
\bibitem [{\citenamefont {Sasaki}\ \emph {et~al.}(2014)\citenamefont {Sasaki},
  \citenamefont {Klein}, \citenamefont {Qiu},\ and\ \citenamefont
  {Sasai}}]{SKQS2014}%
  \BibitemOpen
  \bibfield  {author} {\bibinfo {author} {\bibfnamefont {H.}~\bibnamefont
  {Sasaki}}, \bibinfo {author} {\bibfnamefont {P.}~\bibnamefont {Klein}},
  \bibinfo {author} {\bibfnamefont {B.}~\bibnamefont {Qiu}}, \ and\ \bibinfo
  {author} {\bibfnamefont {Y.}~\bibnamefont {Sasai}},\ }\bibfield  {title}
  {\enquote {\bibinfo {title} {Impact of oceanic-scale interactions on the
  seasonal modulation of ocean dynamics by the atmosphere},}\ }\href@noop {}
  {\bibfield  {journal} {\bibinfo  {journal} {Nat. Commun.}\ }\textbf {\bibinfo
  {volume} {5}},\ \bibinfo {pages} {5636} (\bibinfo {year} {2014})}\BibitemShut
  {NoStop}%
\bibitem [{\citenamefont {Callies}\ \emph {et~al.}(2015)\citenamefont
  {Callies}, \citenamefont {Ferrari}, \citenamefont {Klymak},\ and\
  \citenamefont {Gula}}]{CFKG2015}%
  \BibitemOpen
  \bibfield  {author} {\bibinfo {author} {\bibfnamefont {J.}~\bibnamefont
  {Callies}}, \bibinfo {author} {\bibfnamefont {R.}~\bibnamefont {Ferrari}},
  \bibinfo {author} {\bibfnamefont {J.~M.}\ \bibnamefont {Klymak}}, \ and\
  \bibinfo {author} {\bibfnamefont {J.}~\bibnamefont {Gula}},\ }\bibfield
  {title} {\enquote {\bibinfo {title} {Seasonality in submesoscale
  turbulence},}\ }\href@noop {} {\bibfield  {journal} {\bibinfo  {journal}
  {Nat. Commun.}\ }\textbf {\bibinfo {volume} {6}},\ \bibinfo {pages} {6862}
  (\bibinfo {year} {2015})}\BibitemShut {NoStop}%
\bibitem [{\citenamefont {Qiu}\ \emph {et~al.}(2018)\citenamefont {Qiu},
  \citenamefont {Chen}, \citenamefont {Klein}, \citenamefont {Wang},
  \citenamefont {Torres}, \citenamefont {Fu},\ and\ \citenamefont
  {Menemenlis}}]{Qiu_etal_2018}%
  \BibitemOpen
  \bibfield  {author} {\bibinfo {author} {\bibfnamefont {B.}~\bibnamefont
  {Qiu}}, \bibinfo {author} {\bibfnamefont {S.}~\bibnamefont {Chen}}, \bibinfo
  {author} {\bibfnamefont {P.}~\bibnamefont {Klein}}, \bibinfo {author}
  {\bibfnamefont {J.}~\bibnamefont {Wang}}, \bibinfo {author} {\bibfnamefont
  {H.}~\bibnamefont {Torres}}, \bibinfo {author} {\bibfnamefont {L.~L.}\
  \bibnamefont {Fu}}, \ and\ \bibinfo {author} {\bibfnamefont {D.}~\bibnamefont
  {Menemenlis}},\ }\bibfield  {title} {\enquote {\bibinfo {title} {Seasonality
  in transition scale from balanced to unbalanced motions in the world
  ocean},}\ }\href@noop {} {\bibfield  {journal} {\bibinfo  {journal} {J. Phys.
  Oceanogr.}\ }\textbf {\bibinfo {volume} {48}},\ \bibinfo {pages} {591--605}
  (\bibinfo {year} {2018})}\BibitemShut {NoStop}%
\bibitem [{\citenamefont {Callies}\ \emph {et~al.}(2016)\citenamefont
  {Callies}, \citenamefont {Flierl}, \citenamefont {Ferrari},\ and\
  \citenamefont {Fox-Kemper}}]{CFFF2016}%
  \BibitemOpen
  \bibfield  {author} {\bibinfo {author} {\bibfnamefont {J.}~\bibnamefont
  {Callies}}, \bibinfo {author} {\bibfnamefont {G.}~\bibnamefont {Flierl}},
  \bibinfo {author} {\bibfnamefont {R.}~\bibnamefont {Ferrari}}, \ and\
  \bibinfo {author} {\bibfnamefont {B.}~\bibnamefont {Fox-Kemper}},\ }\bibfield
   {title} {\enquote {\bibinfo {title} {The role of mixed-layer instabilities
  in submesoscale turbulence},}\ }\href@noop {} {\bibfield  {journal} {\bibinfo
   {journal} {J. Fluid Mech.}\ }\textbf {\bibinfo {volume} {788}},\ \bibinfo
  {pages} {5--41} (\bibinfo {year} {2016})}\BibitemShut {NoStop}%
\bibitem [{\citenamefont {Held}\ \emph {et~al.}(1995)\citenamefont {Held},
  \citenamefont {Pierrehumbert}, \citenamefont {Garner},\ and\ \citenamefont
  {Swanson}}]{Held_etal_1995}%
  \BibitemOpen
  \bibfield  {author} {\bibinfo {author} {\bibfnamefont {I.~M.}\ \bibnamefont
  {Held}}, \bibinfo {author} {\bibfnamefont {R.~T.}\ \bibnamefont
  {Pierrehumbert}}, \bibinfo {author} {\bibfnamefont {S.~T.}\ \bibnamefont
  {Garner}}, \ and\ \bibinfo {author} {\bibfnamefont {K.~L.}\ \bibnamefont
  {Swanson}},\ }\bibfield  {title} {\enquote {\bibinfo {title} {Surface
  quasi-geostrophic dynamics},}\ }\href@noop {} {\bibfield  {journal} {\bibinfo
   {journal} {J. Fluid Mech.}\ }\textbf {\bibinfo {volume} {282}},\ \bibinfo
  {pages} {1--20} (\bibinfo {year} {1995})}\BibitemShut {NoStop}%
\bibitem [{\citenamefont {Lapeyre}(2017)}]{Lapeyre2017}%
  \BibitemOpen
  \bibfield  {author} {\bibinfo {author} {\bibfnamefont {G.}~\bibnamefont
  {Lapeyre}},\ }\bibfield  {title} {\enquote {\bibinfo {title} {Surface
  quasi-geostrophy},}\ }\href@noop {} {\bibfield  {journal} {\bibinfo
  {journal} {Fluids}\ }\textbf {\bibinfo {volume} {2}},\ \bibinfo {pages} {7}
  (\bibinfo {year} {2017})}\BibitemShut {NoStop}%
\bibitem [{\citenamefont {Sinha}\ \emph {et~al.}(2019)\citenamefont {Sinha},
  \citenamefont {Balwada}, \citenamefont {Tarshish},\ and\ \citenamefont
  {Abernathey}}]{Sinha_etal_2019}%
  \BibitemOpen
  \bibfield  {author} {\bibinfo {author} {\bibfnamefont {A.}~\bibnamefont
  {Sinha}}, \bibinfo {author} {\bibfnamefont {D.}~\bibnamefont {Balwada}},
  \bibinfo {author} {\bibfnamefont {N.}~\bibnamefont {Tarshish}}, \ and\
  \bibinfo {author} {\bibfnamefont {R.}~\bibnamefont {Abernathey}},\ }\bibfield
   {title} {\enquote {\bibinfo {title} {Modulation of lateral transport by
  submesoscale flows and inertia-gravity waves},}\ }\href@noop {} {\bibfield
  {journal} {\bibinfo  {journal} {J. Adv. Model. Earth Syst.}\ }\textbf
  {\bibinfo {volume} {11}},\ \bibinfo {pages} {1039--1065} (\bibinfo {year}
  {2019})}\BibitemShut {NoStop}%
\bibitem [{\citenamefont {LaCasce}(2008)}]{LaCasce2008}%
  \BibitemOpen
  \bibfield  {author} {\bibinfo {author} {\bibfnamefont {J.~H.}\ \bibnamefont
  {LaCasce}},\ }\bibfield  {title} {\enquote {\bibinfo {title} {Statistics from
  lagrangian observations},}\ }\href@noop {} {\bibfield  {journal} {\bibinfo
  {journal} {Prog. Oceanogr.}\ }\textbf {\bibinfo {volume} {77}},\ \bibinfo
  {pages} {1--29} (\bibinfo {year} {2008})}\BibitemShut {NoStop}%
\bibitem [{\citenamefont {van Sebille}\ \emph {et~al.}(2018)\citenamefont {van
  Sebille}, \citenamefont {Griffies}, \citenamefont {Abernathey}, \citenamefont
  {Adams}, \citenamefont {Berloff}, \citenamefont {Biastoch}, \citenamefont
  {Blanke}, \citenamefont {Chassignet}, \citenamefont {Cheng}, \citenamefont
  {Cotter}, \citenamefont {Deleersnijder}, \citenamefont {{D\"o\"os}},
  \citenamefont {Drake}, \citenamefont {Drijfhout}, \citenamefont {Gary},
  \citenamefont {Heemink}, \citenamefont {Kjellsson}, \citenamefont {Koszalka},
  \citenamefont {Lange}, \citenamefont {Lique}, \citenamefont {MacGilchrist},
  \citenamefont {Marsh}, \citenamefont {{Mayorga Adame}}, \citenamefont
  {McAdam}, \citenamefont {Nencioli}, \citenamefont {Paris}, \citenamefont
  {Piggott}, \citenamefont {Polton}, \citenamefont {{R\"uhs}}, \citenamefont
  {{S. H. A. M. Shah}}, \citenamefont {Thomas}, \citenamefont {Wang},
  \citenamefont {Wolfram}, \citenamefont {Zanna},\ and\ \citenamefont
  {Zika}}]{vanSebille_etal_2018}%
  \BibitemOpen
  \bibfield  {author} {\bibinfo {author} {\bibfnamefont {E.}~\bibnamefont {van
  Sebille}}, \bibinfo {author} {\bibfnamefont {S.~M.}\ \bibnamefont
  {Griffies}}, \bibinfo {author} {\bibfnamefont {R.}~\bibnamefont
  {Abernathey}}, \bibinfo {author} {\bibfnamefont {T.~P.}\ \bibnamefont
  {Adams}}, \bibinfo {author} {\bibfnamefont {P.}~\bibnamefont {Berloff}},
  \bibinfo {author} {\bibfnamefont {A.}~\bibnamefont {Biastoch}}, \bibinfo
  {author} {\bibfnamefont {B.}~\bibnamefont {Blanke}}, \bibinfo {author}
  {\bibfnamefont {E.~P.}\ \bibnamefont {Chassignet}}, \bibinfo {author}
  {\bibfnamefont {Y.}~\bibnamefont {Cheng}}, \bibinfo {author} {\bibfnamefont
  {C.~J.}\ \bibnamefont {Cotter}}, \bibinfo {author} {\bibfnamefont
  {E.}~\bibnamefont {Deleersnijder}}, \bibinfo {author} {\bibfnamefont
  {K.}~\bibnamefont {{D\"o\"os}}}, \bibinfo {author} {\bibfnamefont {H.~F.}\
  \bibnamefont {Drake}}, \bibinfo {author} {\bibfnamefont {S.}~\bibnamefont
  {Drijfhout}}, \bibinfo {author} {\bibfnamefont {S.~F.}\ \bibnamefont {Gary}},
  \bibinfo {author} {\bibfnamefont {A.~W.}\ \bibnamefont {Heemink}}, \bibinfo
  {author} {\bibfnamefont {J.}~\bibnamefont {Kjellsson}}, \bibinfo {author}
  {\bibfnamefont {I.~M.}\ \bibnamefont {Koszalka}}, \bibinfo {author}
  {\bibfnamefont {M.}~\bibnamefont {Lange}}, \bibinfo {author} {\bibfnamefont
  {C.}~\bibnamefont {Lique}}, \bibinfo {author} {\bibfnamefont {G.~A.}\
  \bibnamefont {MacGilchrist}}, \bibinfo {author} {\bibfnamefont
  {R.}~\bibnamefont {Marsh}}, \bibinfo {author} {\bibfnamefont {C.~G.}\
  \bibnamefont {{Mayorga Adame}}}, \bibinfo {author} {\bibfnamefont
  {R.}~\bibnamefont {McAdam}}, \bibinfo {author} {\bibfnamefont
  {F.}~\bibnamefont {Nencioli}}, \bibinfo {author} {\bibfnamefont {C.~B.}\
  \bibnamefont {Paris}}, \bibinfo {author} {\bibfnamefont {M.~D.}\ \bibnamefont
  {Piggott}}, \bibinfo {author} {\bibfnamefont {J.~A.}\ \bibnamefont {Polton}},
  \bibinfo {author} {\bibfnamefont {S.}~\bibnamefont {{R\"uhs}}}, \bibinfo
  {author} {\bibnamefont {{S. H. A. M. Shah}}}, \bibinfo {author}
  {\bibfnamefont {M.~D.}\ \bibnamefont {Thomas}}, \bibinfo {author}
  {\bibfnamefont {J.}~\bibnamefont {Wang}}, \bibinfo {author} {\bibfnamefont
  {P.~J.}\ \bibnamefont {Wolfram}}, \bibinfo {author} {\bibfnamefont
  {L.}~\bibnamefont {Zanna}}, \ and\ \bibinfo {author} {\bibfnamefont {J.~D.}\
  \bibnamefont {Zika}},\ }\bibfield  {title} {\enquote {\bibinfo {title}
  {Lagrangian ocean analysis: Fundamentals and practices},}\ }\href@noop {}
  {\bibfield  {journal} {\bibinfo  {journal} {Ocean Model.}\ }\textbf {\bibinfo
  {volume} {121}},\ \bibinfo {pages} {49--75} (\bibinfo {year}
  {2018})}\BibitemShut {NoStop}%
\bibitem [{\citenamefont {Lumpkin}\ and\ \citenamefont
  {Elipot}(2010)}]{LE2010}%
  \BibitemOpen
  \bibfield  {author} {\bibinfo {author} {\bibfnamefont {R.}~\bibnamefont
  {Lumpkin}}\ and\ \bibinfo {author} {\bibfnamefont {S.}~\bibnamefont
  {Elipot}},\ }\bibfield  {title} {\enquote {\bibinfo {title} {Surface drifter
  pair spreading in the north atlantic},}\ }\href@noop {} {\bibfield  {journal}
  {\bibinfo  {journal} {J. Geophys. Res.}\ }\textbf {\bibinfo {volume} {115}},\
  \bibinfo {pages} {C12017} (\bibinfo {year} {2010})}\BibitemShut {NoStop}%
\bibitem [{\citenamefont {Poje}\ \emph {et~al.}(2014)\citenamefont {Poje},
  \citenamefont {{\"Ozg\"okmen}}, \citenamefont {Jr.}, \citenamefont {Haus},
  \citenamefont {Ryan}, \citenamefont {Haza}, \citenamefont {Jacobs},
  \citenamefont {Reniers}, \citenamefont {Olascoaga}, \citenamefont {Novelli},
  \citenamefont {Griffa}, \citenamefont {Beron-Vera}, \citenamefont {Chen},
  \citenamefont {Coelho}, \citenamefont {Hogan}, \citenamefont {{Kirwan Jr.}},
  \citenamefont {Huntley},\ and\ \citenamefont {Mariano}}]{Poje_etal_2014}%
  \BibitemOpen
  \bibfield  {author} {\bibinfo {author} {\bibfnamefont {A.~C.}\ \bibnamefont
  {Poje}}, \bibinfo {author} {\bibfnamefont {T.~M.}\ \bibnamefont
  {{\"Ozg\"okmen}}}, \bibinfo {author} {\bibfnamefont {B.~L.~L.}\ \bibnamefont
  {Jr.}}, \bibinfo {author} {\bibfnamefont {B.~K.}\ \bibnamefont {Haus}},
  \bibinfo {author} {\bibfnamefont {E.~H.}\ \bibnamefont {Ryan}}, \bibinfo
  {author} {\bibfnamefont {A.~C.}\ \bibnamefont {Haza}}, \bibinfo {author}
  {\bibfnamefont {G.~A.}\ \bibnamefont {Jacobs}}, \bibinfo {author}
  {\bibfnamefont {A.~J. H.~M.}\ \bibnamefont {Reniers}}, \bibinfo {author}
  {\bibfnamefont {M.~J.}\ \bibnamefont {Olascoaga}}, \bibinfo {author}
  {\bibfnamefont {G.}~\bibnamefont {Novelli}}, \bibinfo {author} {\bibfnamefont
  {A.}~\bibnamefont {Griffa}}, \bibinfo {author} {\bibfnamefont {F.~J.}\
  \bibnamefont {Beron-Vera}}, \bibinfo {author} {\bibfnamefont {S.~S.}\
  \bibnamefont {Chen}}, \bibinfo {author} {\bibfnamefont {E.}~\bibnamefont
  {Coelho}}, \bibinfo {author} {\bibfnamefont {P.~J.}\ \bibnamefont {Hogan}},
  \bibinfo {author} {\bibfnamefont {A.~D.}\ \bibnamefont {{Kirwan Jr.}}},
  \bibinfo {author} {\bibfnamefont {H.~S.}\ \bibnamefont {Huntley}}, \ and\
  \bibinfo {author} {\bibfnamefont {A.~J.}\ \bibnamefont {Mariano}},\
  }\bibfield  {title} {\enquote {\bibinfo {title} {Submesoscale dispersion in
  the vicinity of the deepwater horizon spill},}\ }\href@noop {} {\bibfield
  {journal} {\bibinfo  {journal} {Proc. Natl Acad. Sci. USA}\ }\textbf
  {\bibinfo {volume} {111}},\ \bibinfo {pages} {12693–12698} (\bibinfo {year}
  {2014})}\BibitemShut {NoStop}%
\bibitem [{\citenamefont {Poje}\ \emph {et~al.}(2017)\citenamefont {Poje},
  \citenamefont {{\"Ozg\"okmen}}, \citenamefont {Bogucki},\ and\ \citenamefont
  {{Kirwan Jr.}}}]{Poje_etal_2017}%
  \BibitemOpen
  \bibfield  {author} {\bibinfo {author} {\bibfnamefont {A.~C.}\ \bibnamefont
  {Poje}}, \bibinfo {author} {\bibfnamefont {T.~M.}\ \bibnamefont
  {{\"Ozg\"okmen}}}, \bibinfo {author} {\bibfnamefont {D.~J.}\ \bibnamefont
  {Bogucki}}, \ and\ \bibinfo {author} {\bibfnamefont {A.~D.}\ \bibnamefont
  {{Kirwan Jr.}}},\ }\bibfield  {title} {\enquote {\bibinfo {title} {Evidence
  of a forward energy cascade and {Kolmogorov} self-similarity in submesoscale
  ocean surface drifter observations},}\ }\href@noop {} {\bibfield  {journal}
  {\bibinfo  {journal} {Phys. Fluids}\ }\textbf {\bibinfo {volume} {29}},\
  \bibinfo {pages} {020701--020710} (\bibinfo {year} {2017})}\BibitemShut
  {NoStop}%
\bibitem [{\citenamefont {Corrado}\ \emph {et~al.}(2017)\citenamefont
  {Corrado}, \citenamefont {Lacorata}, \citenamefont {Palatella}, \citenamefont
  {Santoleri},\ and\ \citenamefont {Zambianchi}}]{CLPSZ2017}%
  \BibitemOpen
  \bibfield  {author} {\bibinfo {author} {\bibfnamefont {R.}~\bibnamefont
  {Corrado}}, \bibinfo {author} {\bibfnamefont {G.}~\bibnamefont {Lacorata}},
  \bibinfo {author} {\bibfnamefont {L.}~\bibnamefont {Palatella}}, \bibinfo
  {author} {\bibfnamefont {R.}~\bibnamefont {Santoleri}}, \ and\ \bibinfo
  {author} {\bibfnamefont {E.}~\bibnamefont {Zambianchi}},\ }\bibfield  {title}
  {\enquote {\bibinfo {title} {General characteristics of relative dispersion
  in the ocean},}\ }\href@noop {} {\bibfield  {journal} {\bibinfo  {journal}
  {Sci. Rep.}\ }\textbf {\bibinfo {volume} {7}},\ \bibinfo {pages} {46291}
  (\bibinfo {year} {2017})}\BibitemShut {NoStop}%
\bibitem [{\citenamefont {Essink}\ \emph {et~al.}(2019)\citenamefont {Essink},
  \citenamefont {Hormann}, \citenamefont {Centurioni},\ and\ \citenamefont
  {Mahadevan}}]{Essink_etal_2019}%
  \BibitemOpen
  \bibfield  {author} {\bibinfo {author} {\bibfnamefont {S.}~\bibnamefont
  {Essink}}, \bibinfo {author} {\bibfnamefont {V.}~\bibnamefont {Hormann}},
  \bibinfo {author} {\bibfnamefont {L.~R.}\ \bibnamefont {Centurioni}}, \ and\
  \bibinfo {author} {\bibfnamefont {A.}~\bibnamefont {Mahadevan}},\ }\bibfield
  {title} {\enquote {\bibinfo {title} {Can we detect submesoscale motions in
  drifter pair dispersion?}}\ }\href@noop {} {\bibfield  {journal} {\bibinfo
  {journal} {J. Phys. Oceanogr.}\ }\textbf {\bibinfo {volume} {49}},\ \bibinfo
  {pages} {2237--2254} (\bibinfo {year} {2019})}\BibitemShut {NoStop}%
\bibitem [{\citenamefont {Berti}\ \emph {et~al.}(2011)\citenamefont {Berti},
  \citenamefont {Santos}, \citenamefont {Lacorata},\ and\ \citenamefont
  {Vulpiani}}]{BDLV2011}%
  \BibitemOpen
  \bibfield  {author} {\bibinfo {author} {\bibfnamefont {S.}~\bibnamefont
  {Berti}}, \bibinfo {author} {\bibfnamefont {F.~D.}\ \bibnamefont {Santos}},
  \bibinfo {author} {\bibfnamefont {G.}~\bibnamefont {Lacorata}}, \ and\
  \bibinfo {author} {\bibfnamefont {A.}~\bibnamefont {Vulpiani}},\ }\bibfield
  {title} {\enquote {\bibinfo {title} {Lagrangian drifter dispersion in the
  southwestern atlantic ocean},}\ }\href@noop {} {\bibfield  {journal}
  {\bibinfo  {journal} {J. Phys. Oceanogr.}\ }\textbf {\bibinfo {volume}
  {41}},\ \bibinfo {pages} {1659--1672} (\bibinfo {year} {2011})}\BibitemShut
  {NoStop}%
\bibitem [{\citenamefont {Schroeder}\ \emph {et~al.}(2012)\citenamefont
  {Schroeder}, \citenamefont {Chiggiato}, \citenamefont {Haza}, \citenamefont
  {Griffa}, \citenamefont {{\"Ozg\"okmen}}, \citenamefont {Zanasca},
  \citenamefont {Molcard}, \citenamefont {Borghini}, \citenamefont {Poulain},
  \citenamefont {Gerin}, \citenamefont {Zambianchi}, \citenamefont {Falco},\
  and\ \citenamefont {Trees}}]{Schroeder_etal_2012}%
  \BibitemOpen
  \bibfield  {author} {\bibinfo {author} {\bibfnamefont {K.}~\bibnamefont
  {Schroeder}}, \bibinfo {author} {\bibfnamefont {J.}~\bibnamefont
  {Chiggiato}}, \bibinfo {author} {\bibfnamefont {A.~C.}\ \bibnamefont {Haza}},
  \bibinfo {author} {\bibfnamefont {A.}~\bibnamefont {Griffa}}, \bibinfo
  {author} {\bibfnamefont {T.~M.}\ \bibnamefont {{\"Ozg\"okmen}}}, \bibinfo
  {author} {\bibfnamefont {P.}~\bibnamefont {Zanasca}}, \bibinfo {author}
  {\bibfnamefont {A.}~\bibnamefont {Molcard}}, \bibinfo {author} {\bibfnamefont
  {M.}~\bibnamefont {Borghini}}, \bibinfo {author} {\bibfnamefont {P.~M.}\
  \bibnamefont {Poulain}}, \bibinfo {author} {\bibfnamefont {R.}~\bibnamefont
  {Gerin}}, \bibinfo {author} {\bibfnamefont {E.}~\bibnamefont {Zambianchi}},
  \bibinfo {author} {\bibfnamefont {P.}~\bibnamefont {Falco}}, \ and\ \bibinfo
  {author} {\bibfnamefont {C.}~\bibnamefont {Trees}},\ }\bibfield  {title}
  {\enquote {\bibinfo {title} {Targeted lagrangian sampling of submesoscale
  dispersion at a coastal frontal zone},}\ }\href@noop {} {\bibfield  {journal}
  {\bibinfo  {journal} {Geophys. Res. Lett.}\ }\textbf {\bibinfo {volume}
  {39}},\ \bibinfo {pages} {1168} (\bibinfo {year} {2012})}\BibitemShut
  {NoStop}%
\bibitem [{\citenamefont {LaCasce}\ and\ \citenamefont {Bower}(2000)}]{LB2000}%
  \BibitemOpen
  \bibfield  {author} {\bibinfo {author} {\bibfnamefont {J.~H.}\ \bibnamefont
  {LaCasce}}\ and\ \bibinfo {author} {\bibfnamefont {A.}~\bibnamefont
  {Bower}},\ }\bibfield  {title} {\enquote {\bibinfo {title} {Relative
  dispersion in the subsurface north atlantic},}\ }\href@noop {} {\bibfield
  {journal} {\bibinfo  {journal} {J. Mar. Res.}\ }\textbf {\bibinfo {volume}
  {58}},\ \bibinfo {pages} {863--894} (\bibinfo {year} {2000})}\BibitemShut
  {NoStop}%
\bibitem [{\citenamefont {Ollitrault}, \citenamefont {Gabillet},\ and\
  \citenamefont {Verdi\`ere}(2005)}]{OGD2005}%
  \BibitemOpen
  \bibfield  {author} {\bibinfo {author} {\bibfnamefont {M.}~\bibnamefont
  {Ollitrault}}, \bibinfo {author} {\bibfnamefont {C.}~\bibnamefont
  {Gabillet}}, \ and\ \bibinfo {author} {\bibfnamefont {A.~D.}\ \bibnamefont
  {Verdi\`ere}},\ }\bibfield  {title} {\enquote {\bibinfo {title} {Open ocean
  regimes of relative dispersion},}\ }\href@noop {} {\bibfield  {journal}
  {\bibinfo  {journal} {J. Fluid Mech.}\ }\textbf {\bibinfo {volume} {533}},\
  \bibinfo {pages} {381--407} (\bibinfo {year} {2005})}\BibitemShut {NoStop}%
\bibitem [{\citenamefont {Balwada}\ \emph {et~al.}(2020)\citenamefont
  {Balwada}, \citenamefont {LaCasce}, \citenamefont {Speer},\ and\
  \citenamefont {Ferrari}}]{BLSF2019}%
  \BibitemOpen
  \bibfield  {author} {\bibinfo {author} {\bibfnamefont {D.}~\bibnamefont
  {Balwada}}, \bibinfo {author} {\bibfnamefont {J.~H.}\ \bibnamefont
  {LaCasce}}, \bibinfo {author} {\bibfnamefont {K.~G.}\ \bibnamefont {Speer}},
  \ and\ \bibinfo {author} {\bibfnamefont {R.}~\bibnamefont {Ferrari}},\
  }\bibfield  {title} {\enquote {\bibinfo {title} {Relative dispersion in the
  {Antarctic Circumpolar Current}},}\ }\href {\doibase
  https://doi.org/10.1175/JPO-D-19-0243.1} {\bibfield  {journal} {\bibinfo
  {journal} {J. Phys. Oceanogr.}\ } (\bibinfo {year} {2020}),\
  https://doi.org/10.1175/JPO-D-19-0243.1}\BibitemShut {NoStop}%
\bibitem [{\citenamefont {Morrow}\ \emph {et~al.}(2019)\citenamefont {Morrow},
  \citenamefont {Fu}, \citenamefont {Ardhuin}, \citenamefont {Benkiran},
  \citenamefont {Chapron}, \citenamefont {Cosme}, \citenamefont {d'Ovidio},
  \citenamefont {Farrar}, \citenamefont {Gille}, \citenamefont {Lapeyre},
  \citenamefont {Traon}, \citenamefont {Pascual}, \citenamefont {Ponte},
  \citenamefont {Qiu}, \citenamefont {Rascle}, \citenamefont {Ubelmann},
  \citenamefont {Wang},\ and\ \citenamefont {Zaron}}]{Morrow_etal_2019}%
  \BibitemOpen
  \bibfield  {author} {\bibinfo {author} {\bibfnamefont {R.}~\bibnamefont
  {Morrow}}, \bibinfo {author} {\bibfnamefont {L.-L.}\ \bibnamefont {Fu}},
  \bibinfo {author} {\bibfnamefont {F.}~\bibnamefont {Ardhuin}}, \bibinfo
  {author} {\bibfnamefont {M.}~\bibnamefont {Benkiran}}, \bibinfo {author}
  {\bibfnamefont {B.}~\bibnamefont {Chapron}}, \bibinfo {author} {\bibfnamefont
  {E.}~\bibnamefont {Cosme}}, \bibinfo {author} {\bibfnamefont
  {F.}~\bibnamefont {d'Ovidio}}, \bibinfo {author} {\bibfnamefont {J.~T.}\
  \bibnamefont {Farrar}}, \bibinfo {author} {\bibfnamefont {S.~T.}\
  \bibnamefont {Gille}}, \bibinfo {author} {\bibfnamefont {G.}~\bibnamefont
  {Lapeyre}}, \bibinfo {author} {\bibfnamefont {P.-Y.~L.}\ \bibnamefont
  {Traon}}, \bibinfo {author} {\bibfnamefont {A.}~\bibnamefont {Pascual}},
  \bibinfo {author} {\bibfnamefont {A.}~\bibnamefont {Ponte}}, \bibinfo
  {author} {\bibfnamefont {B.}~\bibnamefont {Qiu}}, \bibinfo {author}
  {\bibfnamefont {N.}~\bibnamefont {Rascle}}, \bibinfo {author} {\bibfnamefont
  {C.}~\bibnamefont {Ubelmann}}, \bibinfo {author} {\bibfnamefont
  {J.}~\bibnamefont {Wang}}, \ and\ \bibinfo {author} {\bibfnamefont {E.~D.}\
  \bibnamefont {Zaron}},\ }\bibfield  {title} {\enquote {\bibinfo {title}
  {Global observations of fine-scale ocean surface topography with the surface
  water and ocean topography (swot) mission},}\ }\href@noop {} {\bibfield
  {journal} {\bibinfo  {journal} {Front. Mar. Sci.}\ }\textbf {\bibinfo
  {volume} {6}},\ \bibinfo {pages} {232} (\bibinfo {year} {2019})}\BibitemShut
  {NoStop}%
\bibitem [{\citenamefont {Babiano}\ \emph {et~al.}(1990)\citenamefont
  {Babiano}, \citenamefont {Basdevant}, \citenamefont {Roy},\ and\
  \citenamefont {Sadourny}}]{BBRS1990}%
  \BibitemOpen
  \bibfield  {author} {\bibinfo {author} {\bibfnamefont {A.}~\bibnamefont
  {Babiano}}, \bibinfo {author} {\bibfnamefont {C.}~\bibnamefont {Basdevant}},
  \bibinfo {author} {\bibfnamefont {P.~L.}\ \bibnamefont {Roy}}, \ and\
  \bibinfo {author} {\bibfnamefont {R.}~\bibnamefont {Sadourny}},\ }\bibfield
  {title} {\enquote {\bibinfo {title} {Relative dispersion in two-dimensional
  turbulence},}\ }\href@noop {} {\bibfield  {journal} {\bibinfo  {journal} {J.
  Fluid Mech.}\ }\textbf {\bibinfo {volume} {214}},\ \bibinfo {pages}
  {535--557} (\bibinfo {year} {1990})}\BibitemShut {NoStop}%
\bibitem [{\citenamefont {Foussard}\ \emph {et~al.}(2017)\citenamefont
  {Foussard}, \citenamefont {Berti}, \citenamefont {Perrot},\ and\
  \citenamefont {Lapeyre}}]{FBPL2017}%
  \BibitemOpen
  \bibfield  {author} {\bibinfo {author} {\bibfnamefont {A.}~\bibnamefont
  {Foussard}}, \bibinfo {author} {\bibfnamefont {S.}~\bibnamefont {Berti}},
  \bibinfo {author} {\bibfnamefont {X.}~\bibnamefont {Perrot}}, \ and\ \bibinfo
  {author} {\bibfnamefont {G.}~\bibnamefont {Lapeyre}},\ }\bibfield  {title}
  {\enquote {\bibinfo {title} {Relative dispersion in generalized
  two-dimensional turbulence},}\ }\href@noop {} {\bibfield  {journal} {\bibinfo
   {journal} {J. Fluid Mech.}\ }\textbf {\bibinfo {volume} {821}},\ \bibinfo
  {pages} {358--383} (\bibinfo {year} {2017})}\BibitemShut {NoStop}%
\bibitem [{\citenamefont {Malik}(2018)}]{Malik2018}%
  \BibitemOpen
  \bibfield  {author} {\bibinfo {author} {\bibfnamefont {N.~A.}\ \bibnamefont
  {Malik}},\ }\bibfield  {title} {\enquote {\bibinfo {title} {Turbulent
  particle pair diffusion: A theory based on local and non-local diffusional
  processes.}}\ }\href@noop {} {\bibfield  {journal} {\bibinfo  {journal} {PLoS
  ONE}\ }\textbf {\bibinfo {volume} {13}},\ \bibinfo {pages} {1--29} (\bibinfo
  {year} {2018})}\BibitemShut {NoStop}%
\bibitem [{\citenamefont {Malik}(2019)}]{Malik2019}%
  \BibitemOpen
  \bibfield  {author} {\bibinfo {author} {\bibfnamefont {N.~A.}\ \bibnamefont
  {Malik}},\ }\bibfield  {title} {\enquote {\bibinfo {title} {Turbulent
  particle pair diffusion: Numerical simulations},}\ }\href@noop {} {\bibfield
  {journal} {\bibinfo  {journal} {PLoS ONE}\ }\textbf {\bibinfo {volume}
  {14}},\ \bibinfo {pages} {1--28} (\bibinfo {year} {2019})}\BibitemShut
  {NoStop}%
\bibitem [{\citenamefont {Koszalka}\ \emph {et~al.}(2009)\citenamefont
  {Koszalka}, \citenamefont {Bracco}, \citenamefont {McWilliams},\ and\
  \citenamefont {Provenzale}}]{KBMP2009}%
  \BibitemOpen
  \bibfield  {author} {\bibinfo {author} {\bibfnamefont {I.}~\bibnamefont
  {Koszalka}}, \bibinfo {author} {\bibfnamefont {A.}~\bibnamefont {Bracco}},
  \bibinfo {author} {\bibfnamefont {J.~C.}\ \bibnamefont {McWilliams}}, \ and\
  \bibinfo {author} {\bibfnamefont {A.}~\bibnamefont {Provenzale}},\ }\bibfield
   {title} {\enquote {\bibinfo {title} {Dynamics of wind-forced coherent
  anticyclones in the open ocean},}\ }\href@noop {} {\bibfield  {journal}
  {\bibinfo  {journal} {J. Geophys. Res.}\ }\textbf {\bibinfo {volume} {114}},\
  \bibinfo {pages} {C08011} (\bibinfo {year} {2009})}\BibitemShut {NoStop}%
\bibitem [{\citenamefont {{\"Ozg\"okmen}}\ \emph {et~al.}(2012)\citenamefont
  {{\"Ozg\"okmen}}, \citenamefont {Poje}, \citenamefont {Fischer},
  \citenamefont {Childs}, \citenamefont {Krishnan}, \citenamefont {Garth},
  \citenamefont {Haza},\ and\ \citenamefont {Ryan}}]{Ozgokmen_etal_2012}%
  \BibitemOpen
  \bibfield  {author} {\bibinfo {author} {\bibfnamefont {T.~M.}\ \bibnamefont
  {{\"Ozg\"okmen}}}, \bibinfo {author} {\bibfnamefont {A.~C.}\ \bibnamefont
  {Poje}}, \bibinfo {author} {\bibfnamefont {P.~F.}\ \bibnamefont {Fischer}},
  \bibinfo {author} {\bibfnamefont {H.}~\bibnamefont {Childs}}, \bibinfo
  {author} {\bibfnamefont {H.}~\bibnamefont {Krishnan}}, \bibinfo {author}
  {\bibfnamefont {C.}~\bibnamefont {Garth}}, \bibinfo {author} {\bibfnamefont
  {A.~C.}\ \bibnamefont {Haza}}, \ and\ \bibinfo {author} {\bibfnamefont
  {E.}~\bibnamefont {Ryan}},\ }\bibfield  {title} {\enquote {\bibinfo {title}
  {On multi-scale dispersion under the influence of surface mixed layer
  instabilities and deep flows},}\ }\href@noop {} {\bibfield  {journal}
  {\bibinfo  {journal} {Ocean Model.}\ }\textbf {\bibinfo {volume} {56}},\
  \bibinfo {pages} {16--30} (\bibinfo {year} {2012})}\BibitemShut {NoStop}%
\bibitem [{\citenamefont {Smith}\ \emph {et~al.}(2001)\citenamefont {Smith},
  \citenamefont {Bocaletti}, \citenamefont {Henning}, \citenamefont {Marinov},
  \citenamefont {Tam}, \citenamefont {Held},\ and\ \citenamefont
  {Vallis}}]{Smith_etal_2001}%
  \BibitemOpen
  \bibfield  {author} {\bibinfo {author} {\bibfnamefont {K.~S.}\ \bibnamefont
  {Smith}}, \bibinfo {author} {\bibfnamefont {G.}~\bibnamefont {Bocaletti}},
  \bibinfo {author} {\bibfnamefont {C.~C.}\ \bibnamefont {Henning}}, \bibinfo
  {author} {\bibfnamefont {I.~N.}\ \bibnamefont {Marinov}}, \bibinfo {author}
  {\bibfnamefont {C.~Y.}\ \bibnamefont {Tam}}, \bibinfo {author} {\bibfnamefont
  {I.~M.}\ \bibnamefont {Held}}, \ and\ \bibinfo {author} {\bibfnamefont
  {G.~K.}\ \bibnamefont {Vallis}},\ }\bibfield  {title} {\enquote {\bibinfo
  {title} {Turbulent diffusion in the geostrophic inverse cascade},}\
  }\href@noop {} {\bibfield  {journal} {\bibinfo  {journal} {J. Fluid Mech.}\
  }\textbf {\bibinfo {volume} {469}},\ \bibinfo {pages} {14--47} (\bibinfo
  {year} {2001})}\BibitemShut {NoStop}%
\bibitem [{\citenamefont {Berti}\ and\ \citenamefont
  {Lapeyre}(2014)}]{BL_2014}%
  \BibitemOpen
  \bibfield  {author} {\bibinfo {author} {\bibfnamefont {S.}~\bibnamefont
  {Berti}}\ and\ \bibinfo {author} {\bibfnamefont {G.}~\bibnamefont
  {Lapeyre}},\ }\bibfield  {title} {\enquote {\bibinfo {title} {Lagrangian
  reconstructions of temperature and velocities at submesoscales},}\
  }\href@noop {} {\bibfield  {journal} {\bibinfo  {journal} {Ocean Model.}\
  }\textbf {\bibinfo {volume} {76}},\ \bibinfo {pages} {59--71} (\bibinfo
  {year} {2014})}\BibitemShut {NoStop}%
\bibitem [{\citenamefont {LaCasce}(1996)}]{LaCasce1996}%
  \BibitemOpen
  \bibfield  {author} {\bibinfo {author} {\bibfnamefont {J.~H.}\ \bibnamefont
  {LaCasce}},\ }\emph {\bibinfo {title} {Baroclinic vortices over a sloping
  bottom}},\ \href@noop {} {\bibinfo {type} {{Ph.D.} thesis}},\ \bibinfo
  {school} {MIT/WHOI Joint Program in Physical Oceanography} (\bibinfo {year}
  {1996})\BibitemShut {NoStop}%
\bibitem [{\citenamefont {LaCasce}(1998)}]{LaCasce1998}%
  \BibitemOpen
  \bibfield  {author} {\bibinfo {author} {\bibfnamefont {J.~H.}\ \bibnamefont
  {LaCasce}},\ }\bibfield  {title} {\enquote {\bibinfo {title} {A geostrophic
  vortex on a slope},}\ }\href@noop {} {\bibfield  {journal} {\bibinfo
  {journal} {J. Phys. Oceanogr.}\ }\textbf {\bibinfo {volume} {28}},\ \bibinfo
  {pages} {2362–2381} (\bibinfo {year} {1998})}\BibitemShut {NoStop}%
\bibitem [{\citenamefont {Hua}(1994)}]{Hua1994}%
  \BibitemOpen
  \bibfield  {author} {\bibinfo {author} {\bibfnamefont {B.~L.}\ \bibnamefont
  {Hua}},\ }\bibfield  {title} {\enquote {\bibinfo {title} {The conservation of
  potential vorticity along lagrangian trajectories in simulations of
  eddy-driven flows},}\ }\href@noop {} {\bibfield  {journal} {\bibinfo
  {journal} {J. Phys. Oceanogr.}\ }\textbf {\bibinfo {volume} {24}},\ \bibinfo
  {pages} {498--508} (\bibinfo {year} {1994})}\BibitemShut {NoStop}%
\bibitem [{\citenamefont {Artale}\ \emph {et~al.}(1997)\citenamefont {Artale},
  \citenamefont {Boffetta}, \citenamefont {Celani}, \citenamefont {Cencini},\
  and\ \citenamefont {Vulpiani}}]{ABCCV1997}%
  \BibitemOpen
  \bibfield  {author} {\bibinfo {author} {\bibfnamefont {V.}~\bibnamefont
  {Artale}}, \bibinfo {author} {\bibfnamefont {G.}~\bibnamefont {Boffetta}},
  \bibinfo {author} {\bibfnamefont {A.}~\bibnamefont {Celani}}, \bibinfo
  {author} {\bibfnamefont {M.}~\bibnamefont {Cencini}}, \ and\ \bibinfo
  {author} {\bibfnamefont {A.}~\bibnamefont {Vulpiani}},\ }\bibfield  {title}
  {\enquote {\bibinfo {title} {Dispersion of passive tracers in closed basins:
  beyond the diffusion coefficient},}\ }\href@noop {} {\bibfield  {journal}
  {\bibinfo  {journal} {Phys. Fluids A}\ }\textbf {\bibinfo {volume} {9}},\
  \bibinfo {pages} {3162--3171} (\bibinfo {year} {1997})}\BibitemShut {NoStop}%
\bibitem [{\citenamefont {Aurell}\ \emph {et~al.}(1997)\citenamefont {Aurell},
  \citenamefont {Boffetta}, \citenamefont {Crisanti}, \citenamefont {Paladin},\
  and\ \citenamefont {Vulpiani}}]{ABCPV1997}%
  \BibitemOpen
  \bibfield  {author} {\bibinfo {author} {\bibfnamefont {E.}~\bibnamefont
  {Aurell}}, \bibinfo {author} {\bibfnamefont {G.}~\bibnamefont {Boffetta}},
  \bibinfo {author} {\bibfnamefont {A.}~\bibnamefont {Crisanti}}, \bibinfo
  {author} {\bibfnamefont {G.}~\bibnamefont {Paladin}}, \ and\ \bibinfo
  {author} {\bibfnamefont {A.}~\bibnamefont {Vulpiani}},\ }\bibfield  {title}
  {\enquote {\bibinfo {title} {Predictability in the large: an extension of the
  concept of lyapunov exponent},}\ }\href@noop {} {\bibfield  {journal}
  {\bibinfo  {journal} {J. Phys. A}\ }\textbf {\bibinfo {volume} {30}},\
  \bibinfo {pages} {1--26} (\bibinfo {year} {1997})}\BibitemShut {NoStop}%
\bibitem [{\citenamefont {Cencini}\ and\ \citenamefont
  {Vulpiani}(2013)}]{CV2013}%
  \BibitemOpen
  \bibfield  {author} {\bibinfo {author} {\bibfnamefont {M.}~\bibnamefont
  {Cencini}}\ and\ \bibinfo {author} {\bibfnamefont {A.}~\bibnamefont
  {Vulpiani}},\ }\bibfield  {title} {\enquote {\bibinfo {title} {Finite size
  lyapunov exponent: review on applications},}\ }\href@noop {} {\bibfield
  {journal} {\bibinfo  {journal} {J. Phys. A: Math. Theor.}\ }\textbf {\bibinfo
  {volume} {46}},\ \bibinfo {pages} {254019} (\bibinfo {year}
  {2013})}\BibitemShut {NoStop}%
\bibitem [{\citenamefont {Iudicone}\ \emph {et~al.}(2002)\citenamefont
  {Iudicone}, \citenamefont {Lacorata}, \citenamefont {Rupolo}, \citenamefont
  {Santoleri},\ and\ \citenamefont {Vulpiani}}]{ILRSV2002}%
  \BibitemOpen
  \bibfield  {author} {\bibinfo {author} {\bibfnamefont {D.}~\bibnamefont
  {Iudicone}}, \bibinfo {author} {\bibfnamefont {G.}~\bibnamefont {Lacorata}},
  \bibinfo {author} {\bibfnamefont {V.}~\bibnamefont {Rupolo}}, \bibinfo
  {author} {\bibfnamefont {R.}~\bibnamefont {Santoleri}}, \ and\ \bibinfo
  {author} {\bibfnamefont {A.}~\bibnamefont {Vulpiani}},\ }\bibfield  {title}
  {\enquote {\bibinfo {title} {Sensitivity of numerical tracer trajectories to
  uncertainties in ogcm velocity fields},}\ }\href@noop {} {\bibfield
  {journal} {\bibinfo  {journal} {Ocean Model.}\ }\textbf {\bibinfo {volume}
  {4}},\ \bibinfo {pages} {313--325} (\bibinfo {year} {2002})}\BibitemShut
  {NoStop}%
\bibitem [{\citenamefont {Lacorata}\ \emph {et~al.}(2019)\citenamefont
  {Lacorata}, \citenamefont {Corrado}, \citenamefont {Falcini},\ and\
  \citenamefont {Santoleri}}]{LCFS2019}%
  \BibitemOpen
  \bibfield  {author} {\bibinfo {author} {\bibfnamefont {G.}~\bibnamefont
  {Lacorata}}, \bibinfo {author} {\bibfnamefont {R.}~\bibnamefont {Corrado}},
  \bibinfo {author} {\bibfnamefont {F.}~\bibnamefont {Falcini}}, \ and\
  \bibinfo {author} {\bibfnamefont {R.}~\bibnamefont {Santoleri}},\ }\bibfield
  {title} {\enquote {\bibinfo {title} {Fsle analysis and validation of
  lagrangian simulations based on satellite-derived globcurrent velocity
  data},}\ }\href@noop {} {\bibfield  {journal} {\bibinfo  {journal} {Remote
  Sens. Environ.}\ }\textbf {\bibinfo {volume} {221}},\ \bibinfo {pages}
  {136--143} (\bibinfo {year} {2019})}\BibitemShut {NoStop}%
\bibitem [{\citenamefont {Batchelor}(1950)}]{Batchelor1950}%
  \BibitemOpen
  \bibfield  {author} {\bibinfo {author} {\bibfnamefont {G.~K.}\ \bibnamefont
  {Batchelor}},\ }\bibfield  {title} {\enquote {\bibinfo {title} {The
  application of the similarity theory of turbulence to atmospheric
  diffusion},}\ }\href@noop {} {\bibfield  {journal} {\bibinfo  {journal} {Q.
  J. R. Meteorol. Soc.}\ }\textbf {\bibinfo {volume} {551}},\ \bibinfo {pages}
  {133--146} (\bibinfo {year} {1950})}\BibitemShut {NoStop}%
\bibitem [{\citenamefont {Bourgoin}\ \emph {et~al.}(2006)\citenamefont
  {Bourgoin}, \citenamefont {Ouellette}, \citenamefont {Xu}, \citenamefont
  {Berg},\ and\ \citenamefont {E.Bodenschatz}}]{BOXBB2006}%
  \BibitemOpen
  \bibfield  {author} {\bibinfo {author} {\bibfnamefont {M.}~\bibnamefont
  {Bourgoin}}, \bibinfo {author} {\bibfnamefont {N.~T.}\ \bibnamefont
  {Ouellette}}, \bibinfo {author} {\bibfnamefont {H.}~\bibnamefont {Xu}},
  \bibinfo {author} {\bibfnamefont {J.}~\bibnamefont {Berg}}, \ and\ \bibinfo
  {author} {\bibnamefont {E.Bodenschatz}},\ }\bibfield  {title} {\enquote
  {\bibinfo {title} {The role of pair dispersion in turbulent flow},}\
  }\href@noop {} {\bibfield  {journal} {\bibinfo  {journal} {Science}\ }\textbf
  {\bibinfo {volume} {331}},\ \bibinfo {pages} {835--838} (\bibinfo {year}
  {2006})}\BibitemShut {NoStop}%
\bibitem [{\citenamefont {Spydell}, \citenamefont {Feddersen},\ and\
  \citenamefont {MacMahan}(2020)}]{Spydell_etal_2020}%
  \BibitemOpen
  \bibfield  {author} {\bibinfo {author} {\bibfnamefont {M.~S.}\ \bibnamefont
  {Spydell}}, \bibinfo {author} {\bibfnamefont {F.}~\bibnamefont {Feddersen}},
  \ and\ \bibinfo {author} {\bibfnamefont {J.}~\bibnamefont {MacMahan}},\
  }\bibfield  {title} {\enquote {\bibinfo {title} {Relative dispersion on the
  inner shelf: evidence of a {Batchelor} regime},}\ }\href {\doibase
  https://doi.org/10.1175/JPO-D-20-0170.1} {\bibfield  {journal} {\bibinfo
  {journal} {J. Phys. Oceanogr.}\ } (\bibinfo {year} {2020}),\
  https://doi.org/10.1175/JPO-D-20-0170.1}\BibitemShut {NoStop}%
\bibitem [{\citenamefont {Bourgoin}(2018)}]{Bourgoin2018}%
  \BibitemOpen
  \bibfield  {author} {\bibinfo {author} {\bibfnamefont {M.}~\bibnamefont
  {Bourgoin}},\ }\bibfield  {title} {\enquote {\bibinfo {title} {Some aspects
  of lagrangian dynamics of turbulence},}\ }in\ \href@noop {} {\emph {\bibinfo
  {booktitle} {Mixing and dispersion in flows dominated by rotation and
  buoyancy}}},\ \bibinfo {editor} {edited by\ \bibinfo {editor} {\bibfnamefont
  {H.~J.}\ \bibnamefont {Clercx}}\ and\ \bibinfo {editor} {\bibfnamefont
  {G.~F.~V.}\ \bibnamefont {Heijst}}}\ (\bibinfo  {publisher} {Springer},\
  \bibinfo {year} {2018})\BibitemShut {NoStop}%
\bibitem [{\citenamefont {Boffetta}\ \emph {et~al.}(1999)\citenamefont
  {Boffetta}, \citenamefont {Celani}, \citenamefont {Crisanti},\ and\
  \citenamefont {Vulpiani}}]{BCCV1999}%
  \BibitemOpen
  \bibfield  {author} {\bibinfo {author} {\bibfnamefont {G.}~\bibnamefont
  {Boffetta}}, \bibinfo {author} {\bibfnamefont {A.}~\bibnamefont {Celani}},
  \bibinfo {author} {\bibfnamefont {A.}~\bibnamefont {Crisanti}}, \ and\
  \bibinfo {author} {\bibfnamefont {A.}~\bibnamefont {Vulpiani}},\ }\bibfield
  {title} {\enquote {\bibinfo {title} {Pair dispersion in synthetic fully
  developed turbulence},}\ }\href@noop {} {\bibfield  {journal} {\bibinfo
  {journal} {Phys. Rev. E}\ }\textbf {\bibinfo {volume} {60}},\ \bibinfo
  {pages} {6734} (\bibinfo {year} {1999})}\BibitemShut {NoStop}%
\bibitem [{\citenamefont {Biferale}\ \emph {et~al.}(2005)\citenamefont
  {Biferale}, \citenamefont {Boffetta}, \citenamefont {Celani}, \citenamefont
  {Devenish}, \citenamefont {Lanotte},\ and\ \citenamefont
  {Toschi}}]{BBCDLT2005}%
  \BibitemOpen
  \bibfield  {author} {\bibinfo {author} {\bibfnamefont {L.}~\bibnamefont
  {Biferale}}, \bibinfo {author} {\bibfnamefont {G.}~\bibnamefont {Boffetta}},
  \bibinfo {author} {\bibfnamefont {A.}~\bibnamefont {Celani}}, \bibinfo
  {author} {\bibfnamefont {B.~J.}\ \bibnamefont {Devenish}}, \bibinfo {author}
  {\bibfnamefont {A.}~\bibnamefont {Lanotte}}, \ and\ \bibinfo {author}
  {\bibfnamefont {F.}~\bibnamefont {Toschi}},\ }\bibfield  {title} {\enquote
  {\bibinfo {title} {Lagrangian statistics of particle pairs in homogeneous
  isotropic turbulence},}\ }\href@noop {} {\bibfield  {journal} {\bibinfo
  {journal} {Phys. Fluids}\ }\textbf {\bibinfo {volume} {17}},\ \bibinfo
  {pages} {115101} (\bibinfo {year} {2005})}\BibitemShut {NoStop}%
\bibitem [{\citenamefont {Falkovich}, \citenamefont {Gawedzki},\ and\
  \citenamefont {Vergassola}(2001)}]{FGV2001}%
  \BibitemOpen
  \bibfield  {author} {\bibinfo {author} {\bibfnamefont {G.}~\bibnamefont
  {Falkovich}}, \bibinfo {author} {\bibfnamefont {K.}~\bibnamefont {Gawedzki}},
  \ and\ \bibinfo {author} {\bibfnamefont {M.}~\bibnamefont {Vergassola}},\
  }\bibfield  {title} {\enquote {\bibinfo {title} {Particles and fluids in
  turbulence},}\ }\href@noop {} {\bibfield  {journal} {\bibinfo  {journal}
  {Rev. Mod. Phys.}\ }\textbf {\bibinfo {volume} {73}},\ \bibinfo {pages}
  {913--975} (\bibinfo {year} {2001})}\BibitemShut {NoStop}%
\bibitem [{\citenamefont {Boffetta}\ \emph {et~al.}(2000)\citenamefont
  {Boffetta}, \citenamefont {Celani}, \citenamefont {Cencini}, \citenamefont
  {Lacorata},\ and\ \citenamefont {Vulpiani}}]{BCCLV2000}%
  \BibitemOpen
  \bibfield  {author} {\bibinfo {author} {\bibfnamefont {G.}~\bibnamefont
  {Boffetta}}, \bibinfo {author} {\bibfnamefont {A.}~\bibnamefont {Celani}},
  \bibinfo {author} {\bibfnamefont {M.}~\bibnamefont {Cencini}}, \bibinfo
  {author} {\bibfnamefont {G.}~\bibnamefont {Lacorata}}, \ and\ \bibinfo
  {author} {\bibfnamefont {A.}~\bibnamefont {Vulpiani}},\ }\bibfield  {title}
  {\enquote {\bibinfo {title} {Nonasymptotic properties of transport and
  mixing},}\ }\href@noop {} {\bibfield  {journal} {\bibinfo  {journal} {Chaos}\
  }\textbf {\bibinfo {volume} {10}},\ \bibinfo {pages} {50} (\bibinfo {year}
  {2000})}\BibitemShut {NoStop}%
\bibitem [{\citenamefont {Essink}(2019)}]{Essink2019}%
  \BibitemOpen
  \bibfield  {author} {\bibinfo {author} {\bibfnamefont {S.}~\bibnamefont
  {Essink}},\ }\emph {\bibinfo {title} {Lagrangian dispersion and deformation
  in submesoscale flows}},\ \href@noop {} {\bibinfo {type} {{Ph.D.} thesis}},\
  \bibinfo  {school} {MIT/WHOI Joint Program in Physical Oceanography}
  (\bibinfo {year} {2019})\BibitemShut {NoStop}%
\bibitem [{\citenamefont {Mahadevan}\ \emph {et~al.}(2020)\citenamefont
  {Mahadevan}, \citenamefont {Pascual}, \citenamefont {Rudnick}, \citenamefont
  {Ruiz}, \citenamefont {{Tintor\'e}},\ and\ \citenamefont
  {{D'Asaro}}}]{Mahadevan_etal_2020}%
  \BibitemOpen
  \bibfield  {author} {\bibinfo {author} {\bibfnamefont {A.}~\bibnamefont
  {Mahadevan}}, \bibinfo {author} {\bibfnamefont {A.}~\bibnamefont {Pascual}},
  \bibinfo {author} {\bibfnamefont {D.~L.}\ \bibnamefont {Rudnick}}, \bibinfo
  {author} {\bibfnamefont {S.}~\bibnamefont {Ruiz}}, \bibinfo {author}
  {\bibfnamefont {J.}~\bibnamefont {{Tintor\'e}}}, \ and\ \bibinfo {author}
  {\bibfnamefont {E.}~\bibnamefont {{D'Asaro}}},\ }\bibfield  {title} {\enquote
  {\bibinfo {title} {Coherent pathways for vertical transport from the surface
  ocean to interior},}\ }\href@noop {} {\bibfield  {journal} {\bibinfo
  {journal} {Bull. Am. Meteorol. Soc.}\ }\textbf {\bibinfo {volume} {101}},\
  \bibinfo {pages} {E1996–E2004} (\bibinfo {year} {2020})}\BibitemShut
  {NoStop}%
\end{thebibliography}%

%%%%%%%%%%%%%%%%%%%%%%%%%%%%
%%%%%%%%%%%%%%%%%%%%%%%%%%%%%%%%%%%%%%%%%%%%
\end{document}